# Title: 3D Polycatenated Architected Materials

**Authors:** Wenjie Zhou[1]*, Sujeeka Nadarajah[1], Liuchi Li[2,3], Anna G. Izard[4], Hujie Yan[1], Aashutosh K. Prachet[1], Payal Patel[1], Xiaoxing Xia[4]*, and Chiara Daraio[1]*

**Affiliations:**

[1]Division of Engineering and Applied Science, California Institute of Technology, Pasadena, CA, USA

[2]Department of Civil and Environmental Engineering, Princeton University, Princeton, NJ, USA

[3]Hopkins Extreme Materials Institute, Johns Hopkins University, Baltimore, MD, USA

[4]Lawrence Livermore National Laboratory, Livermore, CA, USA

*Corresponding author. Email: zhouw@caltech.edu, xia7@llnl.gov, daraio@caltech.edu

**Abstract:** Architected materials derive their properties from the geometric arrangement of their internal structural elements. Their designs rely on continuous networks of members to control the global mechanical behavior of the bulk. Here, we introduce a class of materials that consist of discrete concatenated rings or cage particles interlocked in three-dimensional networks, forming polycatenated architected materials (PAMs). We propose a general design framework that translates arbitrary crystalline networks into particles' concatenations and geometries. In response to small external loads, PAMs behave like non-Newtonian fluids, showing both shear-thinning and shear-thickening responses. At larger strains, PAMs behave like lattices and foams, with a nonlinear stress-strain relation. At microscale, we demonstrate that PAMs can change their shapes in response to applied electrostatic charges. PAM's unique properties pave the path for developing stimuli-responsive materials, energy-absorbing systems and morphing architectures.

**Main Text:**

## Introduction

Previously studied architected materials have mostly been designed using rigidly connected truss- (*1*, *2*), plate- (*3*, *4*), or shell-based lattices (*5*, *6*) that derive effective bulk properties from the construct of periodically repeating unit cells (*7*, *8*), or disordered architectures (*9*, *10*). They can display remarkable behaviors such as high strength while being lightweight (*2*, *11*), negative Poisson's ratios (*12*, *13*), and shear-normal coupling (*14*, *15*). Depending on the geometry of their interior structure, they exhibit unusual or extreme mechanical properties (*2*, *11*), such as large reconfigurability (*16–20*), multi-stable responses (*21*, *22*) and nonlinear elastic deformations (*23*, *24*). In some realizations, architected materials behave like fluids – for example, in pentamode materials, which exhibit a near-zero shear response (*25*, *26*). Granular particles and crystals are a type of architected materials that consist of regular arrays of closely packed elements that interact elastically (*27*, *28*). They also present rich mechanical properties, such as geometrical hardening (*29*) and nonlinear wave propagation (*28*), which emerge from the shape of the individual particles, their topological arrangement and the particles' chemical composition. However, in the absence of boundary confinement, granular crystals are not cohesive, because the particles are not held together by binders, and they offer no resistance in tension.

An emergent family of architected materials are topologically interlocked fabrics, consisting of layers of two-dimensional (2D) concatenated particles, like chainmails used in medieval armors, which have been shown to support tunable stiffness and controllable shape morphing (*30*, *31*). Unlike their 2D counterparts, 3D architected materials have many unique deformation modes, which are expected to lead to new materials' behaviors. The vast design space of 3D architected materials offers unprecedented opportunities to tailor mechanical properties. However, to date, there exists no design framework that enables the rational formulation of 3D concatenated structures with designated network topologies.

Here, we integrate the principle of polycatenation (*32–34*) into 3D architected materials that we fabricate at the macro- and micro-scales, to create interlocking arrangements of discrete ring or cage particles building blocks with controllable kinematic degrees of freedoms (DOFs). Such units are symmetrically catenated with their neighbors that are free to move relative to each other within the bound of the interlocking mechanisms. We propose a design strategy to translate continuous graph topologies, such as crystalline lattices (*35*), into their polycatenated counterparts and demonstrate the realization of 3D polycatenated architected materials (PAMs) from selected building blocks. We show how local variations in the particles' geometry affects the internal DOFs, which, in turn, control the global deformability and effective response of the bulk. Because of the discrete nature of the particle-particle interactions in PAMs, their global mechanical behavior transitions from fluid-like to solid-like and from shear-thinning to shear-thickening, as a function of the applied loading. PAMs are resilient to cyclic loading and have tunable energy absorption, with scalable responses that persist at both the macro- and micro-scales. The unique behavior of PAMs arises from strong surface interactions between adjacent particles, for example, in contacts and relative sliding. These interactions can be further leveraged at smaller scales, by increasing the surface-to-volume ratio of the samples. To verify this hypothesis, we fabricate microscale PAMs and show their ability to rapidly and reversibly adapt their shapes in response to electrostatic charges.

## Design approach for PAMs

*Conversion of continuous graph topologies into PAMs*. Traditional lattice structures can be mapped into continuous topological networks that consist of nodes and connections (Fig. 1A). Starting from a chosen crystalline network, we create periodically entangled toroidal, polygonal, or polyhedral caged particles (hereinafter referred to as 'particles') that can be tessellated into PAMs. The process begins by identifying node symmetries in the continuous networks and aligning them with particles that possess these symmetries (Fig. 1B). These particles interlink with adjacent ones, replicating the original network connections (Figs. 1A-C). Thanks to the enormous database of crystallographic symmetries, topological networks from databases like RCSR (*35*) can be transformed into polycatenated analogs (Fig. S1). A single node can be represented using particles with various shapes, such as polyhedral wireframes, polygon clusters, or torus clusters (Fig. 1D). Depending on the nature of the constituent units, the topologies of resulting polycatenated architectures exhibit substantial variations (Fig. 1E).

*Generating 3D PAMs from prescribed particle geometries*. A given particle shape (e.g., cuboctahedron, CO) can exhibit multiple symmetry axes (Fig. 1F), leading to several potential catenation environments (Figs. 1G-I). By utilizing these catenations singularly or in combination, we can create a variety of PAMs (Figs. 1J-M), each with its own global topology. We employ a tripartite naming scheme, **X-n-abc**, for easy identification: 'X' for the network topology (Table S2), 'n' for the number of concatenations per particle, and 'abc' for the particle shape, either in full (lowercase letters) or as an acronym (uppercase letters). As an example, the label **D-4-TET** (Fig. 1C) denotes a polycatenated diamond network (**D**) constituted from tetrahedral (**TET**) wireframes that interlock corner-to-corner with 4 neighboring particles (**4**).

## Mechanical characterization of PAMs

We design $N \times N \times N$ arrays (where $N$ is the number of unit cells consisting of one or more interlocked particles, see Table S1) leading to different PAM geometries and topologies. We fabricate them using additive manufacturing with a brittle acrylic polymer (see Materials and Methods). We selected eight types of representative PAMs, four composed of 2D particles (Figs. S4A-D) and four composed of 3D particles (Figs. S4E-H). After fabrication and removal of support material, the PAMs relaxed under gravity and the originally ordered and periodic microstructure became irregular (Figs. 2A-G). Such gravity-induced shape deformations (hereinafter referred to as 'relaxations') of PAMs are highly dependent on their domain boundaries (e.g., spherical, Fig. 2A-C) and on the particle-to-particle clearance. For mechanical characterization, we designed cubic samples with sides ca. 50 mm (Figs. 2D-G, S4), and measured their designed and relaxed dimensions, as summarized in Table S1.

To characterize their mechanical responses, we conducted quasi-static uniaxial compression, simple shear, and rheology tests under different loading conditions. The mechanical response of PAMs emerges from a complex interplay of interactions across scales, ranging from (i) µm-scale inter-particle contacts, (ii) mm-scale particle deformation (e.g., bending, buckling and fracture), (iii) mesoscale layer-by-layer collapse, and (iv) cm-scale global deformations. Initially, all loadings induce the rearrangement of particles within the available kinematic DOFs and the redistribution of stresses within the volume, without damage to the particles. However, the particles' rearrangement leads to a bulk deformation of PAMs, which persists even after the removal of the external loads. As the particles reach a jammed state, further spatial reconfiguration becomes untenable. Beyond jamming, continued compressive forces result in particles' deformation, damage and fracture.

Under uniaxial compression (Figs. 2H, S6A-F, S7, Movie S2), all samples exhibit a nonlinear stress-strain behavior with significant loading/unloading hysteresis (energy absorption). This hysteric stress-strain relationship is likely influenced by three distinct mechanisms: (i) the rearrangement of catenated particles, (ii) the presence of friction at the contact, and (iii) at larger strains, particle deformation and damage. We note that, after relaxation, the particles' arrangement (i.e., their position and orientation) in PAMs is intrinsically disordered and varies from experiment to experiment. However, we observe that PAMs present consistent values of effective stiffness, within a statistical distribution in the range of 10%.

A particularly interesting observation is the strain-stiffening response during the loading phase (Fig. 2J), which is reminiscent of the response of classical granular systems (*36*). To understand and quantify the role of dynamic contact chains within PAMs, we perform level-set discrete element method (LS-DEM) (*37*, *38*) simulations with particles of arbitrary shapes (*30*, *39*). We first analyze the response of a T-6-ring sample, subjected to uniaxial compression (see Methods). Experimentally, under cyclic compression at lower strains, PAMs show a reduced peak stress and a hysteretic response with a progressively smaller area, then stabilize into a steady-state response after the first few cycles (Fig. S7).

Within the small to moderate strain regime (e.g., no particle fracture), our numerical simulation quantitatively captures the steady-state stress-strain response (see Fig. 2J for an example). From the simulation result, we further calculate the apparent compressive modulus (E*) at each loading step by fitting a polynomial to the stress-strain curve during the loading phase (Fig. 2J, inset). As the compression progresses, we observe the T-6-ring PAM "densifies" as more contacts (both tensile and compressive ones) form among particles (Fig. 2K) and carry larger contact forces (thicker and longer blue and red lines in Fig. 2K). To relate the macroscale strain-stiffening response with the particle-scale "densification" process, we plot E* as a function of $\langle Z \rangle - Z_0$, where $\langle Z \rangle$ is the mean contact number (summing over cohesive and compressive ones) and $Z_0$ is the minimal contact number per particle for the sample to become structurally rigid (see Methods in SM) (*40*). $Z_0$ depends on many factors such as particle friction, geometry, and catenation topology and is unknown for these catenated granular systems. Nevertheless, we find picking $Z_0 = 5$ can collapse data from different preparation protocols and loading strain ranges (Fig. 2L), which follows a power-law scaling as predicted by the critical phenomenon of jamming phase transition (*40*). Here, the scaling exponent is very close to one, obtained by fitting a second-order polynomial to the stress-strain curve to get E*. This result, together with additional particle-scale analysis based on granular physics (see Fig. S11), suggests that the (small-strain) mechanics of PAMs share many features with that of classical granular materials. Note that, however, these LS-DEM simulation-based analyses will fail when particles deform or fracture.

Unlike other architected materials, PAMs exhibit a notable ability to adjust their interparticle arrangements in response to external loads, a characteristic also found in granular systems. This particles' rearrangement leads to two mechanical regimes, which appear under all deformation modes: (i) a fluid-like response, with a vanishing shear modulus, linked to relative particles' motion; and (ii) a solid-like response, characterized by particles' deformation, beyond the jamming transition. To describe the fluid-like mechanical response, we conducted simple shear and rheology tests on J-4-ring and T-6-ring samples (Figs. 3A,B). These PAMs were selected because they are composed of toroidal particles, which have greater kinematic DOFs. To control testing conditions, we fabricated samples that incorporated top and bottom gripping plates (Figs. 3A,B, S5).

Under shear loads, both J-4-ring and T-6-ring samples demonstrated a plateau region with force values close to zero, indicative of a fluid-like behavior (Figs. 3C,F, Movie S3). Beyond a critical strain, such plateau then transitioned to a quasi-linear elastic region, typical of solid-like behavior. This transition can be correlated to the reduced DOFs between rings, which jam under external tensile, compressive, or shear loads. The fluid-like and solid-like regimes can be programmed by designing the catenation topologies and particle geometries.

To further understand this fluid-solid duality, we characterized the rheological properties (technical details in supplementary text) of cylindrical shaped PAM samples (Fig. 3B, Movie S4). In oscillatory amplitude sweep experiments (Figs. 3D,G), the J-4-ring and T-6-ring samples initially displayed a decrease in storage modulus (G'), loss modulus (G"), and complex viscosity ($\eta^*$) with increasing torsional strain. Upon reaching the critical jamming strain observed in simple shear tests, all three parameters begin to increase (*41*). In the oscillatory frequency sweep experiments (Figs. 3E,H), both J-4-ring and T-6-ring samples exhibited an unusual inflection in viscosity at high angular frequency. When subjected to torsional strains below their respective jamming transition thresholds, a notable transition from shear-thinning to shear-thickening was observed with increasing oscillation angular frequency. In the latter part of the shear-thickening phase, the substantial increase in viscosity is likely influenced by the inertia effects of particles under high-frequency oscillation conditions. Although both shear-thinning and shear-thickening behaviors have been observed in various materials (*41*, *42*), they have not been reported concurrently in the same material, particularly with the observed pronounced reduction in viscosity followed by a substantial increase, as a function of both angular frequency or torsional strain.

To better understand this unusual frequency-dependent thinning-to-thickening transition, we modeled the response of PAMs using the LS-DEM. We focus on modeling the rheological experiments, and the inflection of $\eta^*$ with increasing angular frequency. Since the mechanical deformation of each particle is minimal compared to the translation observed in their rigid body motion (e.g., rearrangement) in the experiments, we model the rings as rigid particles. Here, we use LS-DEM to model sweep experiments for both J-4-ring and T-6-ring samples. We construct digital twins, replicating the ring's shape, density, size, and spatial arrangement, as well as the total number of rings in each respective sample.

All simulations qualitatively capture the inflection of $\eta^*$ observed in experiments (compare Figs. 3I,K with Figs. 3E,H). Our simulations also agree with experiments in that the value of $\eta^*$ at the inflection point is smaller for the J-4-ring sample compared to that for the T-6-ring sample. However, the simulations overestimate the angular frequency at which the inflection of $\eta^*$ happens. This mismatch could be due to imprecisions in the 3D printed particle geometry, contacts' imperfections, and the presence of friction, which are not included in our models. These discrepancies can also lead to considerable differences in the sample's packing structure in the relaxed configuration in experiments and simulations, which can shift the angular frequency at which the inflection of $\eta^*$ happens. Nevertheless, our simulations provide particle-scale details that allow us to better understand the mechanisms underpinning this thinning-to-thickening transition. More specifically, from the point view of the rheophysics of dense granular materials (*43–45*), $\eta^*$ takes contribution from two components: a "contact" component (which corresponds to percolating and enduring force chains in the statics of granular media, used often in the soil mechanics community), and a "fluctuation" component (which corresponds to the degree of turbulency of granular flow stemming from short-lived particle collisions, used often in the fluid mechanics community).

In our experiments, as the excitation frequency increases, it is expected that both samples experience a transition from a contact-dominant (or "quasi-static") regime to a fluctuation-dominant (or "inertia") regime. As such, the initial decrease of $\eta^*$ may be understood as the decrease of contact (due to stronger centrifugal effects) in the contact-dominant regime, while the later increase of $\eta^*$ can be understood as the increased degree of "turbulency" (due to a faster external excitations) in the fluctuation-dominant regime. This concept is confirmed by looking at Fig. 3I (inset), which shows, for the simulation of the J-4-ring sample, the variation of the average contact number per particle, and of the normalized average particle velocity fluctuation (computed as the square root of *granular temperature* (*43*), see Materials and Methods for our calculation procedure), as functions of the angular frequency. As the angular frequency increases, the average contact number per particle decreases while the particle velocity fluctuation increases. In particular, the latter shows a much rapid increase rate once exceeding the frequency of the inflection point of $\eta^*$, which suggests the transition from the contact-dominant regime to the fluctuation-dominant regime. A similar observation can be made for the T-6-ring sample, by looking at Fig. 3K (inset) and comparing it to Fig. 3I (inset). According to kinetic theory (*46*), the viscosity of granular flow, $\eta$ , depends on the density and granular temperature through a scaling $\eta \sim (\rho_c - \rho)^{-\alpha} \delta v$, where $\rho_c$ is the material density at random close packing, $\rho$ is the material density at a given flow state, and $\alpha$ is a scaling exponent whose value depends on particle properties (such as shape and surface friction). We test the relevance of this theory in describing the rheological response of PAMs. Due to difficulties in calculating objectively $\rho$ for our simulated PAM samples, we use the mean contact number $\langle Z \rangle$ in place of $\rho$ (with $Z_c$ the contact number at random close packing), assuming a power-law scaling between $\rho$ and $\langle Z \rangle$ can translate from conventional granular materials (*44*) to PAMs. Fig. 3J and Fig. 3L show the variation of the normalized viscosity, $\frac{\eta^*}{\delta v}$, as a function of $Z_c - \langle Z \rangle$ for the J-4-ring sample and the T-6-ring sample, respectively. We observe that in the low excitation frequency regime (before the inflection of $\eta^*$), the rheological response follows a power-law scaling (black dashed line in Figs. 3J,L) as predicted by kinetic theory. However, as progressing into the high frequency domain (after the inflection of $\eta^*$), the rheological response deviates further away from the respective power-law scaling (insets of Figs. 3J,L). The exact reason for this deviation is unknown, but one possible cause is the breakdown of the power lawing scaling between $\langle Z \rangle$ and $\rho$ in the high frequency domain due to the presence of tensile contact (for preventing particles from separating) that is absent in conventional granular materials. Lastly, we note that, $Z_c$ is unknown for PAMs. For results shown in Figs. 3J,L, we use the mean contact number of the sample before torsion experiment (that is, after gravity settlement), as an estimation of $Z_c$. A different $Z_c$ will change the scaling exponent presented in Fig. 3J,L, but it will not change the observed power-law relation.

## Programmable critical jamming strains

We study the role of particle geometry and particle's linking topology in the jamming transition in PAMs. The local catenation topologies in PAMs play a pivotal role in defining their mesoscale (e.g., cell-to-cell, layer-by-layer) DOFs, which in turn dictates the critical strain for jamming ($\varepsilon_c^*$:

critical compressive jamming strain; $\gamma_s^*$: critical shear jamming strain). For instance, J-4-ring PAMs show a significantly higher $\gamma_s^*$ as compared to T-6-ring PAMs, whereas T-6-ring exhibits a higher $\varepsilon_c^*$ relative to J-4-ring. In J-4-ring PAMs, shear loading induces a coordinated rearrangement of particles, thereby amplifying $\gamma_s^*$(Fig. 4A). During this process, rings oriented parallel to the shearing direction maintain their orientations, while those perpendicular to the shearing direction rotate in a coordinated manner, facilitating a greater $\gamma_s^*$. However, under compression or tension, the rings' inability to adjust their positions – restricted by limited DOFs from neighboring particles – results in a reduced $\varepsilon_c^*$. In contrast, T-6-ring PAMs rely on a 'scissor' mechanism (Fig. 4C): Upon compressive loading, all rings adjust their orientations coordinatively, which allows for a larger $\varepsilon_c^*$.

In addition to the choice of catenation topology, the thickness, *d,* of the torus significantly influences the jamming transition in PAMs (Fig. 4D). A decrease in *d* generally correlates with an increased DOF in PAMs. To elucidate this relationship, we fabricated a series of J-4-ring and T-6-ring PAMs with constant ring diameters (*D*) but varied *d*. We measured the critical jamming strain under both shear ($\gamma_s^*$) and compressive ($\varepsilon_c^*$) loading conditions as a function of thickness *d* (Fig. 4B). Regardless of the catenation topology, reductions in *d* are associated with increases in both $\gamma_s^*$ and $\varepsilon_c^*$. Furthermore, we observed that catenation topology significantly impacts the predominant deformation modes. For instance, the $\gamma_s^*$ for J-4-ring PAM (Fig. 4D), which exhibit the largest *d/D* ratio and therefore the lowest DOF, is higher compared to T-6-ring PAMs (Fig. 4O), which have a much lower *d/D* ratio.

**Length-scale dependence and electrostatic reconfiguration of μ-PAMs**

The deformability of PAMs is primarily influenced by their particle geometries and the particle-to-particle DOF, which are expected to be scale independent. Hence, we hypothesize that reducing the particle size by a factor of ca. 60 will retain the characteristic mechanical response of macroscopic systems, such as their quasistatic compressive behavior and liquid-fluid duality. To validate this hypothesis, we fabricate PAMs using two-photon lithography with post-printing oxygen plasma etching (see Materials and Methods). Upon completion of the fabrication and release process—where plasma etching removes thin support materials required during fabrication – the μ-PAMs demonstrated a gravitational relaxation analogous to that observed in their macroscopic counterparts (Fig. S17). We designed a series of C-6-TT PAMs (as in Fig. S9) varying their volume fractions, by changing the beam thicknesses of the particles. Following an increasing order of beam thickness, we label the PAMs as I, II, and III (Fig. S9). The same PAMs were fabricated at both macroscale (Ω_I, Ω_II, Ω_III) and microscale (μ_I, μ_II, μ_III), scaling them by a factor of 60 in all dimensions (i.e., sample side lengths: 24 mm and 400 μm). Due to the differences in fabrication methods, we used slightly different acrylic polymers (Material and Methods) for the macro- and micro-scale samples. Nevertheless, we found qualitative agreement between the mechanical responses of PAMs across scales (Figs. 5A-C). The energy absorption capacities of all C-6-TT PAMs were calculated by integrating the areas under the stress-strain curves. Our experiments reveal that scale factors ($U_\mu$/ $U_\Omega$) among all I, II, III designs to be near constant of 12.76±0.53 (Fig. 5A).

One defining difference between micro-scale and macro-scale PAMs is their dramatically different surface-to-volume ratio (~60 times larger in the μ-PAM samples) and the reduced weight of each particle in the catenated network (~216,000 times lower in the μ-PAM samples). Such discrepancy can be exploited to observe the role of inter-particle forces (e.g., electrostatic repulsion) in the global deformation of PAMs. To test this hypothesis, we tessellated J-4-ring μ-PAM voxels to form various geometries: a side-anchored cube (Fig. 5D), a point-anchored cube (Fig. 5E), a bottom-anchored numeral '1' (Fig. 5F), and a bottom-anchored letter 'T' (Fig. 5G). We then coat each μ-PAM sample with a thin layer of copper, approximately 300 nm in thickness, to provide electrical conductivity. We then positioned the samples atop a Van de Graaff generator with direct electrical contact (Fig. S17). As electrostatic charges accumulate, the individual rings within the μ-PAMs began to repel each other, due to increased electrostatic repulsion. This electrostatic interaction prompted the μ-PAMs to both expand outward in all directions due to inter-ring repulsion and elongate upwards due to repulsion between the μ-PAMs and the substrate against gravity, transforming each initially collapsed structure into a structurally deployed state (Figs. 5D-G, Movie S6). The charged μ-PAMs will stay in this deployed geometry until we discharge the Van de Graaff generator by neutralizing the electrical charge stored in its metallic dome, to which the conductive substrate of the μ-PAM samples is attached. The transition between the uncharged, compact state and the charged, deployed state was fast (<0.1 s, Movie S9) and completely reversible. This behavior is derived from the fluid-solid duality also observed in the macroscale samples. Global deformations are allowed by the intrinsic DOF, but constrained when reaching the tensile critical jamming strains, resulting in a 'lockable' 3D shape-morphing behavior. This suggests that μ-PAMs driven by electrostatic forces can be engineered as responsive elements in remotely actuated materials, for micro-scale devices and smart material systems.

## Discussions

Our work introduces PAMs as a novel class of architected materials, distinguished by their ability to exhibit both fluid-like and solid-like behaviors regulated by a jamming transition, which we demonstrate both at the macro- and micro-scales. The generalizable design strategy rationally maps discrete, topologically interlocked particles onto 3D crystalline networks, providing a foundational framework for designing and studying a library of these building blocks. This approach allows PAMs to exhibit tunable mechanical properties, which can be precisely controlled by the geometry and arrangement of the particles. Their nonlinear elastic response in the jammed state, coupled with non-Newtonian shear-thinning and shear-thickening behaviors in the unjammed state, provides a versatile mechanical platform for applications requiring adaptive stiffness and energy dissipation. This is particularly relevant for stimuli-responsive materials, soft robotics, and morphing architectures. PAMs effectively bridge the gap between discrete granular materials and continuous architected materials within a 3D domain.

**Acknowledgments:** We thank Prof. Melany L. Hunt, Prof. Ruby Xiaojing Fu, Dr. Tingtao Zhou, and Jagannadh Boddapati for discussions.

**Funding:** W.Z. and C.D. acknowledge support from the Gary Clinard Innovation fund, and the Army Research Office (MURI ARO W911NF-22-2-0109). Computational resources were provided by the High-Performance Computing Center at Caltech. A.G.I and X.X. acknowledges the financial support from Lawrence Livermore National Laboratory's Lab Directed Research and Development Program (22-ERD-004). Work at LLNL was performed under the auspices of the U.S. Department of Energy by Lawrence Livermore National Laboratory under Contract DE-AC52-07NA27344.

**Author contributions:** W.Z. and S.N. contributed equally. W.Z. conceived the idea, designed the structures and fabricated the samples. W.Z., S.N., and C.D. designed the experiments. S.N., W.Z., A.K.P., and P.P. performed the experiments and analyzed experimental data. L.L. and H.Y. performed numerical simulations and analyzed simulation data. X.X. and W.Z designed the microscale experiments. X.X. and A.G.I fabricated and tested microscale samples. W.Z. and C.D. wrote the initial draft. All authors interpreted the results and reviewed the manuscript.

**Competing interests:** Authors declare that they have no competing interests.

**Data and materials availability:** All data are available in the main text or the supplementary materials. Other information related to this study are available from the corresponding author upon reasonable request.

**Supplementary Materials**

Materials and Methods

Supplementary Text

Figs. S1 to S23

Tables S1 to S2

Reference (*47-58*)

Movies S1 to S9

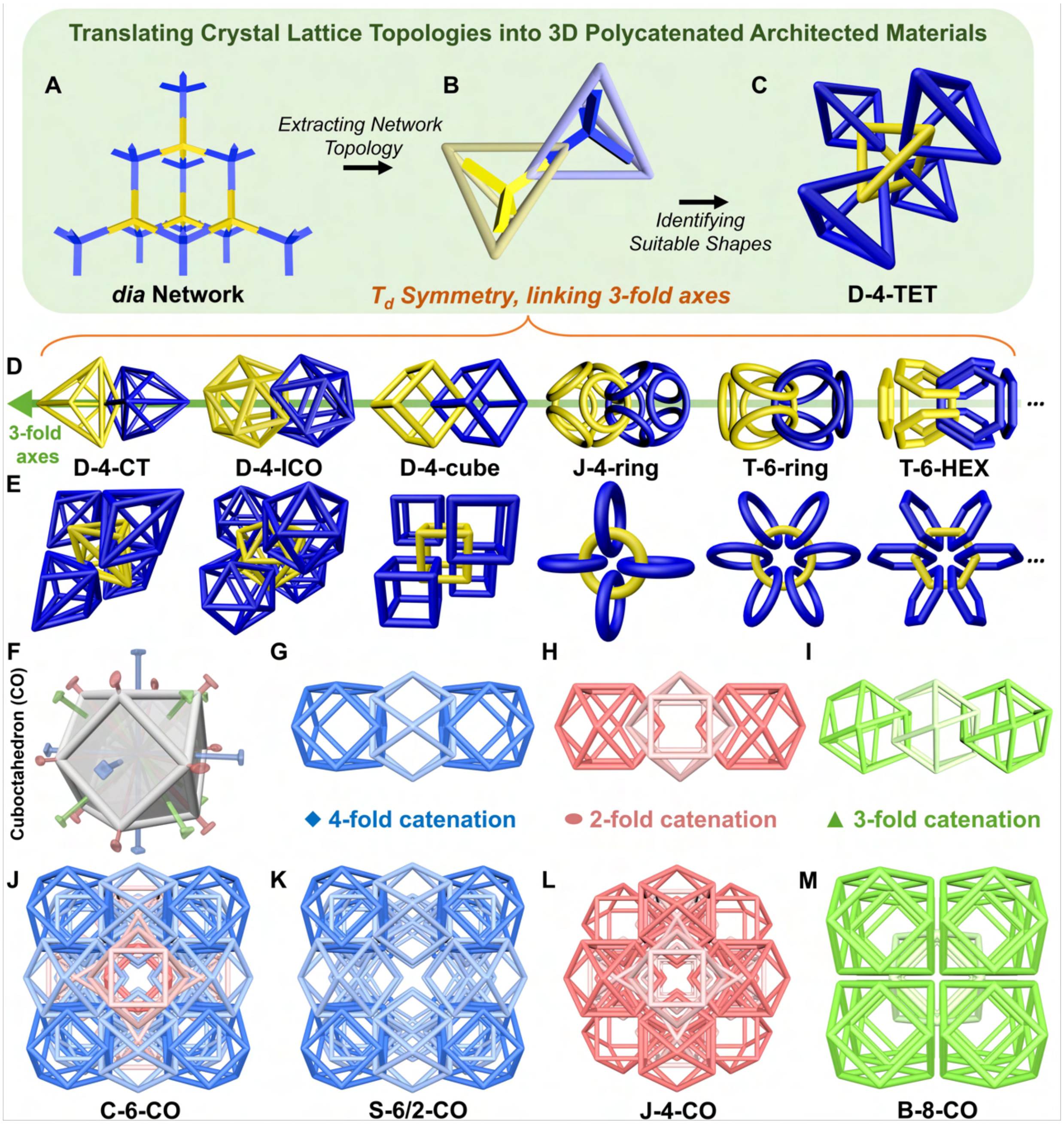

**Fig. 1. Design strategy for PAMs.** (A-E) A typical design workflow of PAMs from a designated network: (A) Network of the *dia* topology. (B) Essential nodes and their connections with adjacent nodes in the *dia* network. Each pair of nodes are mapped with two tetrahedral particles catenating their vertices, while aligning their 3-fold axes. (C) An extended D-4-TET PAM composed of tetrahedral particles. (D) An array of exemplary PAM variants in addition to the catenated tetrahedra, namely Catalan tetrahedra (D-4-CT), icosahedra (D-4-ICO), cubes (D-4-cube), octahedrally arranged six-ring clusters (J-4-ring), tetrahedrally arranged four-ring clusters (T-6-

ring), and tetrahedrally arranged four-hexagon clusters (T-6-HEX). The green line highlights the 3-fold symmetrical axis. (E) A series of extended PAMs corresponding to the configurations from (D). (F-M) Generation of PAMs from designated particle geometry: (F) A cuboctahedron (CO), catenated through its 4-fold (blue), 3-fold (green), and 2-fold (red) axes. (J-M) Expanded PAMs from the catenations illustrated in (G-I).

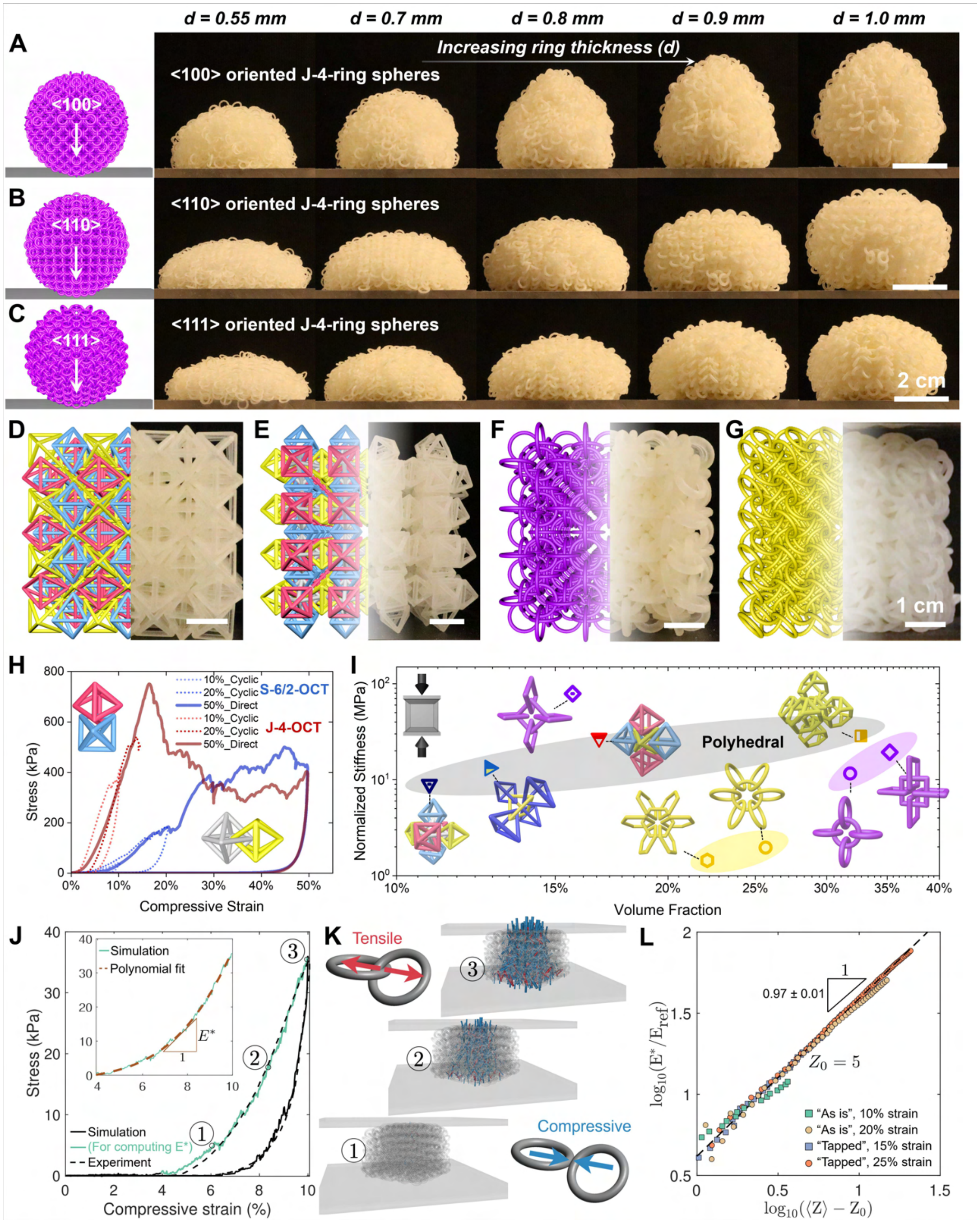

**Fig. 2. Gravity-induced relaxation and uniaxial compression of PAMs.** (A-C) Relaxation of a series of J-4-ring PAMs with spherical domain boundaries, placed on a flat surface. When oriented along the <100> (A), <110> (B), and <111> (C) crystallographic axes, they show different relaxed outline shapes. Scale bar: 2 cm. (D-G) Illustration and corresponding photos of expanded J-4-OCT, S-6/2-OCT, J-4-ring, and T-6-ring PAM samples. Scale bars: 1 cm. (H) Stress-strain plots of J-4-OCT and S-6/2-OCT PAMs under uniaxial compression. (I) Summary of normalized stiffness against measured volume fractions at 5%~10% strain for each PAM. (J) A comparison of the response of a T-6-ring sample under uniaxial compression between experiment and LS-DEM simulation result. The inset shows the calculation of E* by fitting a polynomial to the simulated stress-strain response during the loading phase. (K) A visualization of the contact forces (blue being compressive ones and red being tensile ones) at three selected loading stages (see J). Each contact force is represented by a cylinder, whose orientation is the force vector direction and whose length and radius are scaled according to the force magnitude. (L) Relation between E* and $\langle Z \rangle - Z_0$ for four simulations considering different pre-loading configurations and loading strain ranges. All data can be reasonably represented by a power-law scaling (black dashed line) by setting $Z_0 = 5$.

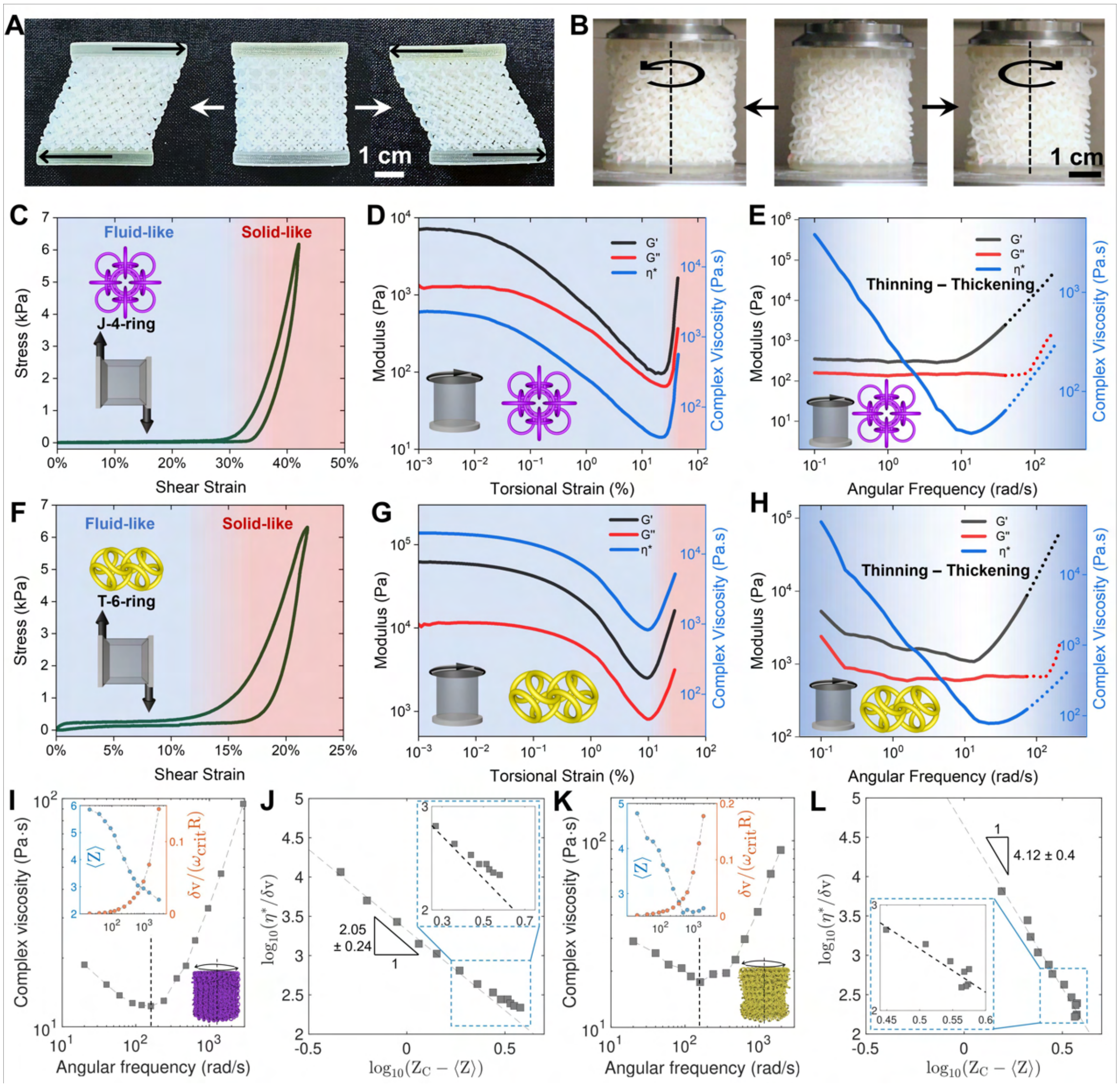

**Fig. 3. Shear and rheology test of PAMs.** (A, B) PAM samples in shearing and rheological test, showing the deformation in response to corresponding loads. (C, F) Stress-shear strain results for J-4-ring (C) and T-6-ring (F) PAMs, showing the transition from fluid-like to solid-like regime with the increase of shear strain. (D, E, G, H) Rheology results of J-4-ring (D) and T-6-ring (G) PAMs under oscillatory amplitude sweep (E) and frequency sweep (H), with plots of storage modulus (G'), loss modulus (G"), and complex viscosity ($\eta^*$) as a function of torsional strain and angular frequency. Dotted lines indicate the region where the tests were significantly affected by the instrument inertia effects. The blue shaded area in (C, D, F, G) indicates the region where the PAMs exhibit fluid-like behavior, transitioning to solid-like behaviors as indicated by the red shaded area. (I, K) Simulation results of the variation of $\eta^*$ as a function of angular frequency for J-4-ring (I) and T-6-ring (K) cylindrical samples. Insets show the mean contact number, $\langle Z \rangle$, (left axis) and normalized granular temperature, $\frac{\delta v}{\omega_{crit} R}$ (right axis) as a function of angular frequency. Here, $\omega_{crit}$ is the angular frequency at which the inflection of $\eta^*$ occurs in our simulations and R is the radii of the sample, used only for making $\delta v$ dimensionless. (J, L) Simulation results of the variation of $\frac{\eta^*}{\delta v}$ as a function of $Z_C - Z$ for the J-4-ring (I) and T-6-ring(K) cylindrical samples. The black dashed lines are power-law fits using simulation data before the inflection of $\eta^*$. Insets show a magnified visualization of the main plot including a portion of the data after the inflection of $\eta^*$.

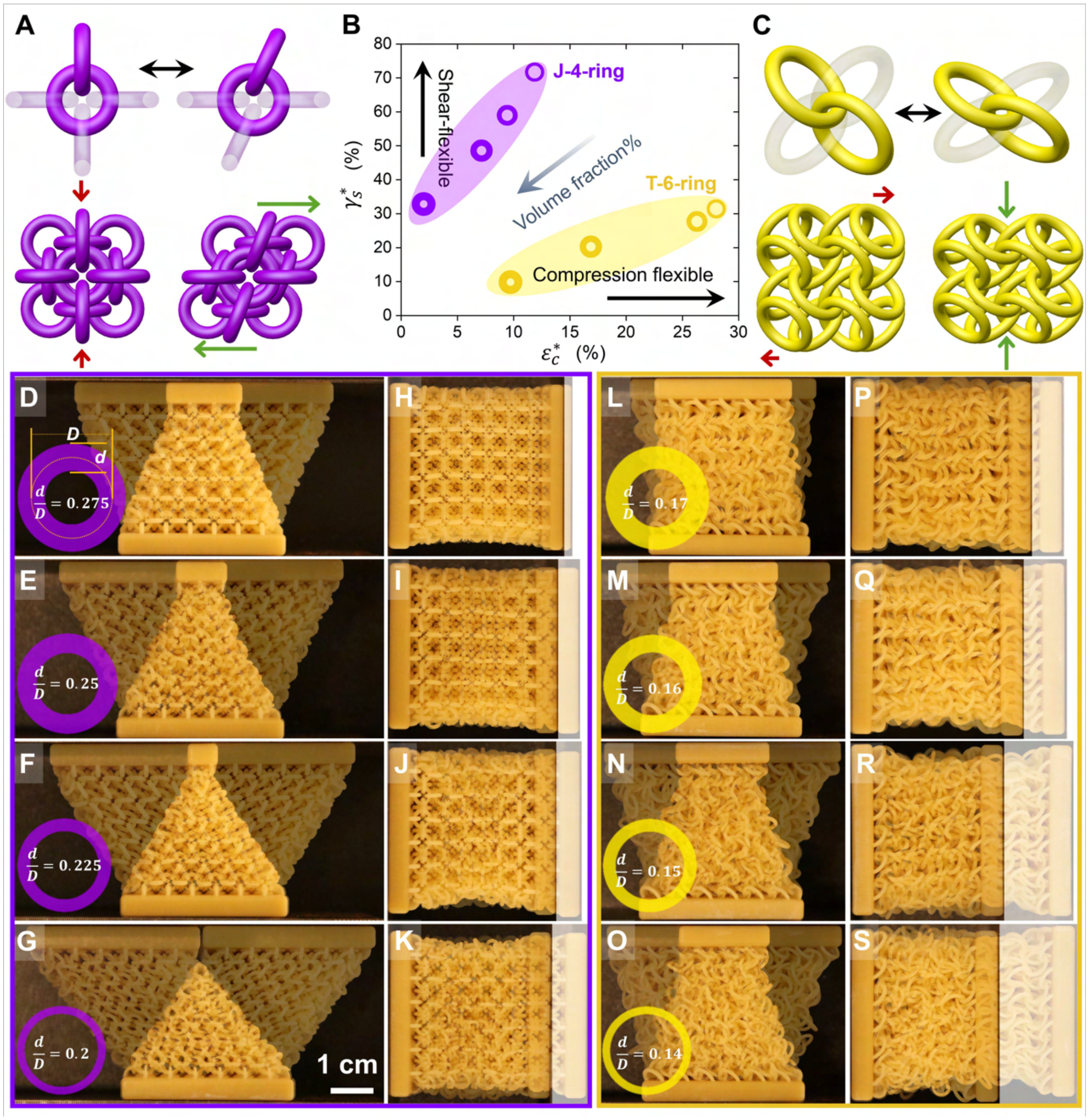

**Fig. 4. Programmable critical jamming strains of PAMs.** (A, C) The schematics illustrate the local ring arrangements under shearing and compressive loads for J-4-ring and T-6-ring PAMs, respectively. (B) A summary plot displaying critical shear jamming strain ($\gamma_s^*$) against the critical compressive jamming strain ($\varepsilon_c^*$) for both types of PAMs containing rings with varied thicknesses. (D-S) Photographs showing J-4-ring (D-K) and T-6-ring (L-S) PAMs at their corresponding critical jamming strains.

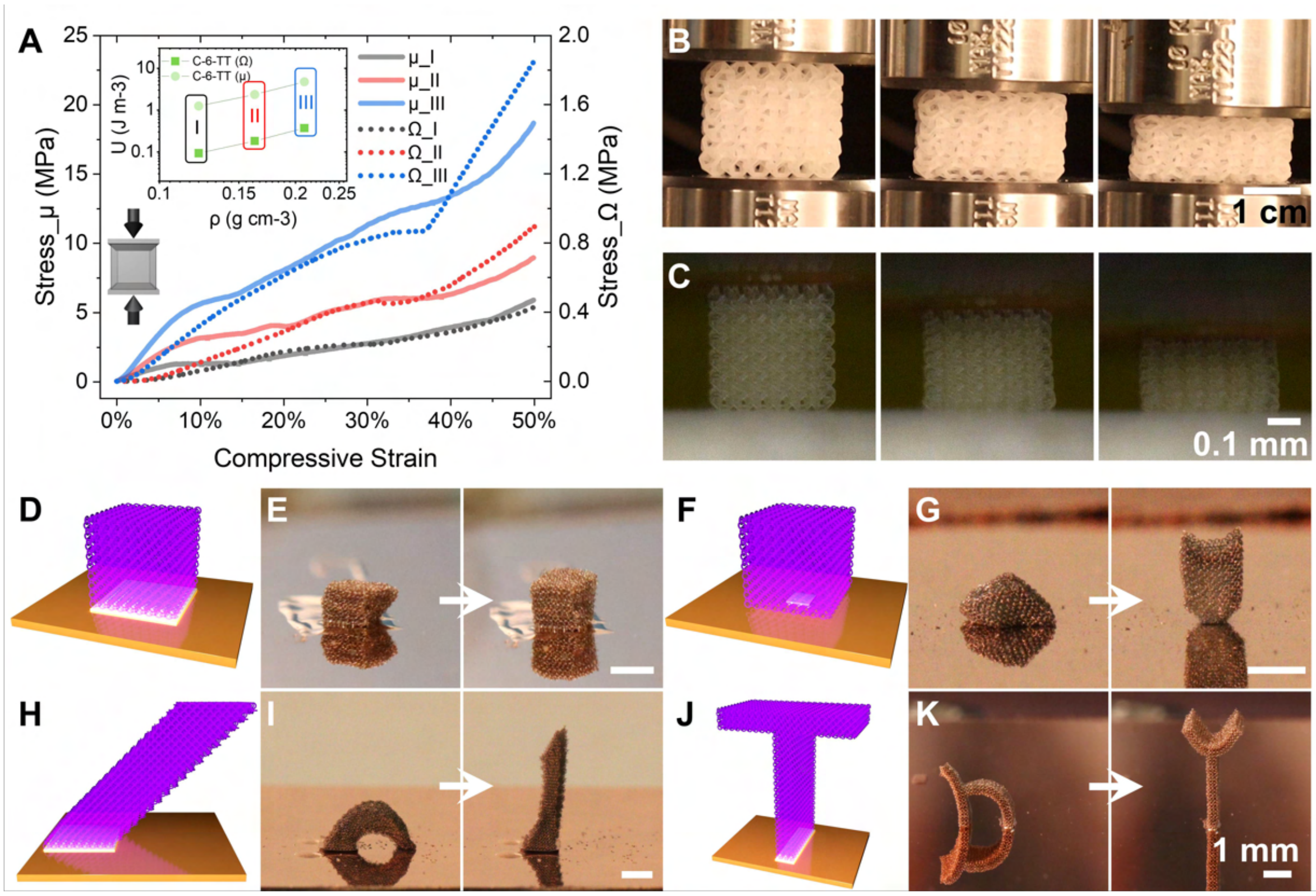

**Fig. 5. Scale independence of PAMs and electrostatic actuation of μ-PAMs.** (A) Stress-strain curves of C-6-TT PAMs fabricated at different scales (60 times difference in all dimensions) and volume fractions (III>II>I). Inset shows a comparative summary of energy absorption capacities of PAMs at two scales. (B, C) Snapshots of macroscale and microscale (B) C-6-TT PAMs undergoing compression experiments at strains of 0%, 25%, and 50%. Scale bars: 1 cm (B), 0.1 mm (C). (D-G) The J-4-ring μ-PAMs in various geometries including cubes and letters before and after electrostatic expansion. White regions in illustrations highlight the fixing area between the μ-PAMs and the substrate. These μ-PAMs, when subjected to electrostatic charges generated by a Van de Graaff generator, deploy from a relaxed natural state to an expanded state due to inter-particle electrostatic repulsion. Scale bars: 1 mm.

# Supplementary Materials for

## 3D Polycatenated Architected Materials

Wenjie Zhou*, Sujeeka Nadarajah, Liuchi Li, Anna G. Izard, Hujie Yan, Aashutosh K. Prachet, Payal Patel, Xiaoxing Xia*, and Chiara Daraio*

*Corresponding authors: zhouw@caltech.edu, xia7@llnl.gov, daraio@caltech.edu

**The PDF file includes:**

Materials and Methods
Supplementary Text
Figs. S1 to S23
Tables S1 to S2
References (*47-58*)

**Other Supplementary Materials for this manuscript include the following:**

Movies S1 to S9

## Materials and Methods

Materials and Fabrication of Macroscopic PAMs

In this study, we fabricated eight representative PAMs, four composed of 2D particles (Figs S4E-H) and four composed of 3D particles (Figs. S4M-P), using a brittle acrylic polymer using additive manufacturing. Specifically, PAM samples were fabricated from 3D Systems employing the Multi Jet Fusion fabrication process, sourced from a commercial 3D printing service (Shapeways). Materials used in this study are Visijet M3, M2R-Clear, and M2R-TN. All mechanical tests were conducted using PAMs fabricated using Visijet M3. Visijet M3 has a density of 1.02 g/cm$^3$, a tensile modulus of 1463 MPa, and a tensile strength of 42.4 MPa (*47*). Upon receiving the samples, some had remaining supporting wax materials, which could (i) alter the particle geometries and (ii) limit particles' DOFs. These wax materials can be selectively dissolved by soaking the samples in isopropanol solution for 2 min at a temperature between 40 and 60 °C. All PAM samples were manually checked to ensure no remaining particle agglutination before mechanical characterizations. To test PAM samples under shear loading (in simple shear and rheology tests), we incorporated two solid gripping plates in the sample's design and fabrication (Fig. S5). The plates partially overlapped with the particles at the top and bottom surfaces and were used to mount the samples to the loading grips and distribute the load uniformly across all particles on the surfaces.

Notably, PAMs can also be fabricated using different materials such as Nylon (selective laser sintering, SLS), TPU (SLS), or metal (selective laser melting, SLM) (Fig. S19). Each sample was designed to be a cube with side ca. 50 mm (Figs. S4E-H, 2M-P). To maintain 0.1 mm particle-particle separation, wax was used as support materials and later dissolved.

Structural Characteristics of PAMs in This Study

PAMs are characterized by their catenation number (CN), defined as the number of neighboring particles that are catenated with the center one, represented using ball-and-stick models (*48*) (Figs. S4A-D, 4SI-L). The same CN can represent multiple local particle arrangements. For example, CN = 4 in a planar arrangement maps to a network with a **nbo** topology, while CN = 4 in a tetrahedral ($T_d$) arrangement maps to a network with a **dia** topology (*35*).

J-4-ring (Fig. S4A) and J-4-square (Fig. S4B) were designed to compare the influence of particle geometry while keeping the PAM topology constant. T-6-ring (Fig. S4C) and T-6-HEX (Fig. S4D) served as another pair of comparison. J-4-OCT (Fig. S4I) and S-6/2-OCT (Fig. S4J) were designed to compare the influence of PAM topology while keeping the particle geometry constant. D-4-TET (Fig. S4K) and C-6-TT (Fig. S4L) were designed because they are two results of geometric transformations from T-6 topology – D-4-TET emerge when every two tetrahedral clusters catenated through 3-fold axes are transformed into a pair of tetrahedra; while C-6-TT emerge when every two tetrahedral clusters catenated through 2-fold axes are transformed into a pair of truncated tetrahedra (Fig. S2). Notably, J-4-square has two stable configurations (Movie S7): an expanded, stiff configuration (L); and a collapsed, flexible configuration (S). Configuration L can transition into S via an auxetic mechanism like the 'rotating squares behavior' (*49*) (Movie S8), which can be trained by cyclic application of external loading. To minimize structural randomness, we standardized the maximum permissible particle thicknesses to maintain clearances as 0.1 mm in all designs except for J-4-OCT and S-6/2-OCT.

Quasi-static Compression and Shear Tests

Quasi-static uniaxial compression and simple shear experiments were conducted using an Instron ElectroPuls E3000 universal testing machine. This testing machine was equipped with an ATI Mini85 multi-axial Force/Torque sensor capable of measuring both force and torque along three mutually perpendicular axes. The load cell's capacity was 1.9 kN and 3.8 kN for force measurements in the X, Y, and Z directions respectively, and 80 Nm for torque measurement along all three axes. For the quasi-static compression trials, cubic specimens were carefully positioned between the compression platens. The orientation ensured that the centerline of the specimen was aligned with the centerline of the load cell. To mount the shear test specimens in the Instron machine, a pair of custom-designed mounting plates was affixed to the solid gripping plates (3D-printed alongside the specimen). The specimens were oriented such that the solid gripping plates aligned with the sides of the specimen. Each mounting plate was subsequently connected to the corresponding upper and lower parts of the tensile grip in the Instron. This modification transformed the uniaxial tensile/compression test apparatus into a configuration suitable for simple shear testing. The force-displacement responses of all specimens were recorded during quasi-static compression tests, at a displacement-controlled loading rate of 0.1/s, to maintain experimental consistency. In the compression tests, a series of multiple compression cycles were performed on a single specimen, each with a varying maximum loading strain spanning from 10% to 50%. Due to the inherently stochastic nature of PAMs, their dimensions exhibited variability in each instance, influenced by random internal arrangement patterns. Thus, dimensions of relaxed PAMs under gravity were measured multiple times, across various regions. The resultant average measurements were then utilized to compute effective stress and strain, accounting for this variability.

Numerical simulation of uniaxial compression experiments (LS-DEM)

Numerical uniaxial compression experiments were conducted using LS-DEM. The model parameters, as well as the construction of digital twins, follows those reported in a previous work (*30*) where LS-DEM was utilized to study the tunable mechanics of structured fabrics. The inter-particle friction between the ring particles was set as 0.25, same as in the previous work (*30*). The inter-particle friction between the two platens and the ring particles was set to a small value of 0.1 (*39*), considering the smoothness of the loading frame in the experiments.

The T-6-ring sample was first settled under gravity to equilibrium and then compressed by moving the top platen downward. The imposed displacement of the top platen, as well as the contact force exerted by particles to the top platen was recorded at each time step, and these two quantities were used to plot the stress-strain result. Note that the dimensions (height, width, and length) of the sample were calculated at the end of the gravity-settling step and then used to convert displacement into strain and contact force into stress.

The apparent compressive modulus (E*) of the sample at each strain value was calculated by fitting a polynomial to the stress-strain curve. For all simulated cases, we consider data starting from 4% strain and onward, excluding the initial stage (< 4% strain) where the stress is close to zero. The observation that E* scales as a power law with $Z - Z_0$ was found to be insensitive to the degree of fitting polynomial used (tested up to degree of five), and the scaling exponent was found to stay around 0.9 (see Fig. S10). $Z_0 = 5$ is picked approximately as the mean contact number of the "tapped" sample before undergoing uniaxial compression. (A "tapped" sample is one further subjected to a small-strain (5% in this work) loading and unloading cycle after settles to equilibrium under gravity, while a "as is" sample is one subjected to only settlement under gravity.) With $Z_0 = 5$ fixed, we also find the scaling exponent value being insensitive to the different order of the polynomial we used for fitting the stress-strain curve (and for subsequently calculating E*, see Fig. S10). Hence, we may regard the "tapped" configuration as being representative of the "marginally jammed" configuration, in terms of finding $Z_0$. Changing $Z_0$ to 5 (obtained from the initial configuration of the "tapped" case) from 2 (obtained from the initial configuration of the "as is" case) was able to collapse data from the "as is" cases onto the data of "tapped cases". This observation suggests that the initial configuration of "tapped" may be used to determine the minimal contact number for the considered sample arrangement to gain structural rigidity.

Fig. S11 show two additional particle-scale analysis of the "as is" 10% strain simulation case. The mean "engaged" friction between ring particles was computed as a function of cumulative strain (Fig. S11a). The Gini coefficient of the magnitude of contact forces was also computed as a function of cumulative strain (Fig. S11b), following the formula presented in (*50*). The Gini coefficient has a value between 1 (totally heterogeneous) and 0 (totally homogeneous) and it is used here to quantify the heterogeneity of contact forces. It was observed that, first, during the loading phase, the engaged friction decreases, and the contact forces become less heterogeneous (the Gini coefficient decreases); second, after entering the unloading phase, the engaged friction increases, and the contact forces become more heterogeneous. Both observations agree with an experimental study (*50*) on the contact force distribution evolution within a micro glass sphere assembly under uniaxial compression.

Rheology Tests

The dynamic oscillatory rheological tests were performed using a torque-controlled Discovery HR20 Hybrid Rheometer, from TA Instruments which has axial force and torque limits of 50 N and 0.2 Nm with sensitivities at 0.005 N and $10^{-10}$ Nm, respectively. The experiment was carried out using a 40 mm diameter parallel plate tool, maintaining a constant temperature of 25 °C and a frequency of 0.5 Hz, with the torsional strain ranging from $10^{-6}$ % to 100%. Cylindrical specimens of original height and diameter of 40 mm were employed. These specimens were designed with thin solid components at both ends, strategically incorporated to ensure steadfast attachment to the rheometer and prevent any slippage. To achieve this secure attachment, thin, robust double-sided adhesive tapes were utilized. Torsional rotation was applied to the specimen through the upper parallel plate, while the bottom side remained securely fixed. Data acquisition was facilitated using the TRIOS software. Data points lying within the strain range of $10^{-6}$ % to $10^{-3}$ %, were subjected to noise interference and were deliberately excluded from the study.

Numerical Simulation of rheological experiments (LS-DEM)

Numerical frequency sweep experiments were conducted using LS-DEM. The model parameters, as well as the construction of digital twins, follows those reported in a previous work (*30*) where LS-DEM was utilized to study the tunable mechanics of structured fabrics. The inter-particle friction for the J-4-ring was set as 0.25, same as in the previous work(*30*). The inter-particle friction for the T-6-ring sample was increased to 0.5, to partially account for the noticeable surface roughness of the 3D-printed ring particles.

Each numerical experiment consists of two steps: settling under gravity until equilibrium and rotating under a given angular frequency. Note that, rings in the bottom layer were fixed during both steps, while those in the top layer were fixed during the first step and were rotated with respect to the sample's vertical central axis during the second step. The rotation in the second step was imposed according to an applied sinusoidal shear strain $\gamma(t) = A\sin(\omega t)$, where $A$ is the amplitude (set at 0.1 for all simulations) and ω is the angular frequency. By doing so, we resemble as close as possible the experimental loading conditions. In practice, the rotation angle $\theta(t)$ and the speed $\dot{\theta}(t)$ were calculated, from which the spatial position, orientation, as well as translational and angular velocity of each ring in the top layer (which revolves around the rotation axis) can be determined. $\theta(t)$ and $\dot{\theta}(t)$ were calculated according to the following two expressions:

$$\theta(t) = \frac{\gamma H}{R}\text{Asin}(\omega t),$$

$$\dot{\theta}(t) = \frac{\gamma H}{R}\text{A}\omega\text{cos}(\omega t).$$

Above, $H$ is the height and $R$ is the radius of the cylindrical sample, determined from the sample's settled configuration. As rings in the top layer were moved via the imposed motion described above, the torque they felted, $\mathrm{T}(t)$, was computed. This is done by summing the cross products between the positions of them (with respect to the rotation axis) and the contact forces (projected onto the plane perpendicular to the rotation axis) felted by them. The shear stress $\tau(t)$ was then computed according to the following expression:

$$\tau(t) = \frac{\mathrm{T}(t)R}{J},$$

where $J = \frac{\pi R^4}{2}$ is the polar moment of inertia of the sample (treated as a rigid cylinder). The maximum value $\tau_{\max}$ was then determined from $\tau(t)$ and was used to find the complex viscosity $\eta^*$ based on the following expression:

$$\eta^* = \frac{\tau_{\max}}{A\omega}.$$

For each simulation with a given angular frequency, we can get multiple values of $\tau_{\max}$ (see Fig. S12 for a representative example), thereby calculating multiple values for $\eta^*$. These values were averaged, and the mean values were reported in the main text. Variations of $\eta^*$ due to variations in $\tau_{\max}$ were found to be small (one order of magnitude smaller) compared to the mean value of $\eta^*$ and are therefore not reported in the main text. For this work, the number of $\tau_{\max}$ value varied from two to ten depends on the imposed angular frequency and out of consideration of computational costs. In general, the larger the angular frequency, the cheaper (in terms of computational cost) the simulation becomes for getting more values for $\tau_{\max}$.

At each location where $\tau_{\max}$ is computed, we also compute the mean contact number per particle $\langle Z\rangle$ and the mean particle velocity fluctuation $\delta v$ by averaging all particles in a sample. The former is easy to get directly from our LS-DEM simulation outputs. For the latter, we take the particle velocity output and perform a spatial average to get $\delta v$. Here, we consider $\delta v$ along the azimuth direction only, and we partition particles into different layers vertically ($z_i$, with $z$ being the height coordinate location) and different annular bins ($r_i$, with $r$ being the radial coordinate location) within each layer. Then, $\delta v$ is computed as:

$$\delta v = \frac{1}{N_z}\sum_{i=1}^{N_z}\frac{1}{N_r}\sum_{j=1}^{N_r}\langle\delta v\rangle_{z_i r_j},$$

where $N_z$ and $N_r$ is the number of layers and bins per layer, respectively. $\langle\delta v\rangle_{z_i r_j}$ is the particle-averaged velocity fluctuation associated with a particular bin in a given layer, and it is computed as the root square of the granular temperature $\langle T\rangle_{z_i r_j}$:

$$\langle\delta v\rangle_{z_i r_j} = \sqrt{\langle T\rangle_{z_i r_j}},$$

with $\langle T\rangle_{z_i r_j} = \frac{1}{N_{p,z_i r_j}}\sum_{k=1}^{N_{p,z_i r_j}}\left(v_{k,r} - \langle v_{k,r}\rangle\right)^2.$

Above, $N_{p,z_i r_j}$ is the number of particles residing in that considered bin. Within this bin, $v_{k,r}$ is the projected velocity along the azimuth direction of the k-th particle:

$$v_{k,r} = \overrightarrow{v_k}\cdot\overrightarrow{t_k},$$

where $\overrightarrow{v_k}$ is the velocity vector of the k-th particle and $\overrightarrow{t_k}$ is the tangential unit vector along the azimuth direction associated with that particle. Lastly, $\langle v_{k,r} \rangle$ is the mean projected velocity, computed as the following:

$\langle v_{k,r} \rangle = \frac{1}{N_p} \sum_{l=1}^{N_p} v_{k,r}^{(l)}$, where we drop the $z_i$ and $r_j$ subscripts for simplicity.

Numerical Simulation of Critical Jamming Experiments

The critical jamming simulations were conducted using the LS-DEM framework. Model parameter settings, including particle representation and inter-particle interactions, follow the rheological simulations previously presented. In the shearing tests, the rings in the bottom layer were fixed, while a constant horizontal force was applied to the top layer in both leftward and rightward directions until the shearing strain reached a steady state. Shearing strains were recorded for both directions, and their mean value was taken. For the compressive straining tests, the rings in the left layer were fixed, and a constant horizontal force was applied to the right layer. Initially, a rightward force was applied to stretch the model to its maximum length, followed by a leftward force to compress the model to its minimum length. The average strain, calculated from both the stretched and compressed states, was reported (Fig. S13). Each simulation for a given configuration and loading condition was repeated multiple times, with the average of the resulting strains presented in Fig. S14.

Fabrication and Characterization of Microscopic PAMs

The μ-PAM samples were fabricated by two-photon lithography using a commercial 3D printer (Nanoscribe GT2) and a commercial resin (Nanoscribe IP-S). The smaller samples used in micro-mechanical compression experiments were printed as designed, without adding additional mechanical supports. The larger samples used in electrostatic deployment experiments were printed by stitching multiple segments sequentially due to the limited field of view of the 25X objective used for two-photon lithography. To prevent drifting issues of the interlocked but unsupported structures due to the prolonged printing time for larger μ-PAM structures, we incorporated additional support structures (body-centered cubic lattices with lateral beam thickness of 500 nm) that overlapped with the μ-PAM lattices in the CAD design. The combined geometry was printed simultaneously so that the body-centered cubic scaffold provided additional support and confinement of the PAM during the printing process. The as-printed samples were immersed in propylene glycol methyl ether acetate (Sigma-Aldrich) for 1hr and then in isopropanol for 1hr to rinse off the remaining uncured liquid resin. The cleaned samples were dried in air. The larger samples used in electrostatic deployment experiments were etched in a PIE Scientific Tergeo Plus Plasma Cleaner for 10-20 h (direct mode, oxygen flow rate 10 sccm, Ar flow rate 30 sccm, RF power 30 W, pulse duty cycle 50/255), to remove the supporting scaffold. Special attention was paid to increase uniformity of etching for all polymer surfaces so that the 500 nm thick body-centered cubic supporting lattices could be removed before substantially damaging the μ-PAM structure. A Keyence VHX-7000 Digital Microscope was used to inspect the etched μ-PAM samples every hour to complete removal of the supporting lattice and release of the interlocked particles in the polycatenated architecture.

For electrostatic deployment experiments, the fully released μ-PAM samples attached to ITO-coated glass substrates were coated with nominally 300 nm thick of Cu (measured on a planar substrate) using a Kurt Lesker PRO Line PVD 75 thin film sputtering system at 100 W DC power. The Cu-coated samples were attached to a Lethan Corporation Van De Graaff Generator using double-sided conductive Cu tapes. Digital photos and videos were taken when the Van De Graaff generator was being turned on and off.

For microscale mechanical testing experiments, the μ-PAMs were fabricated on 5 mm-by-5 mm Si substrates directly without the additional supporting scaffold. The uniaxial compression tests were conducted using a displacement-controlled micromechanical testing system (Kamrath Weiss Tensile & Compression Module). The device was equipped with custom compression grips, including an alumina flat punch with a tip of 2 mm in diameter, a load cell of 10 N, and a 3D-printed fixture to facilitate accurate and convenient mounting of the 5 mm Si substrates onto the compression module. Samples were tested at a strain rate of 0.5%/s. The experiments were performed under a Keyence VHX-7000 digital microscope for in situ video recording.

## Supplementary Text

Structural Turnabilities of PAMs

By changing the particles' shapes within similar symmetries (like from octahedra to cuboctahedra) within a given topology (Fig. S2, Movie S6), we can alter the locking mechanisms (e.g., transitioning from a corner-to-corner linking, as in a J-4-OCT to an edge-to-edge interlocking in the J-4-CO), thus varying the DOFs between particles (Fig. S2). Reducing the particles' thickness, thereby decreasing the volume fraction, typically increases the kinematic DOFs between units within the same global topology and the resulting mechanical response of the PAMs.

To explore a broader property space of PAMs, we identify three orthogonal tuning methods: volume fraction tuning, geometry tuning, and topology tuning. Each method allows for precise control over the mechanical properties and responses of PAMs, enabling a wide range of potential applications.

*Volume Fraction Tuning.* The mechanical characteristics of PAMs, such as critical jamming strains and moduli, can be significantly influenced by adjusting the volume fraction of the materials. This is achieved by varying the beam thicknesses ($d$) of the particles, whether they are polyhedral, polygonal, or torus-shaped. Increasing the beam thickness results in a higher volume fraction, which typically leads to smaller critical jamming strains (Fig. 4B) and larger moduli (Fig. 5A).

*Particle Geometry Tuning.* For a given crystalline network topology, an infinite variety of particle geometries can be designed. For instance, torus particles can be replaced with polygonal particles (Figs. S4A-D), resulting in similar (Figs. S6A,B) or distinct (Figs. S6D,E) mechanical behaviors. Additionally, polyhedral particles can be partially truncated while maintaining their symmetries, altering the local interlinking mechanisms among corner-to-corner, face-to-face, and edge-to-edge locking (Fig. S2). By fine-tuning these kinematic degrees of freedom (DOFs), we can achieve vast turnability in both solid-like and fluid-like mechanical properties.

*Catenation Topology Tuning.* The mapping-based design strategy allows for flexible adaptation of catenation topologies. Either singular or multiple particles can be mapped onto a node in a network topology, resulting in varied catenation structures (Fig. 1D). This flexibility enables the fine-tuning of catenation topologies by substituting a single polyhedral particle with a cluster of spatially arranged planar particles (Fig. S2), provided they share the same symmetry. Such substitutions increase the kinematic DOFs of the resulting PAMs, as planar particles can move relative to each other. Consequently, these PAMs can exhibit drastically different mechanical behaviors, enhancing their adaptability and performance across various applications.

Collectively, these tuning methods empower the design of PAMs with tailored mechanical properties, paving the way for innovative solutions in stimuli-responsive materials, energy-absorbing systems, and morphing architectures.

PAMs' Behavior under Uniaxial Compression

Repeated loading and unloading cause PAMs to develop a permanent residual strain. Under cyclic compression at lower strains, PAMs show a reduced peak stress and a hysteretic response with a progressively smaller area, leading to less energy being dissipated in each cycle (Fig. S7). Most of this reduction happens within the first few cycles, after which the PAMs stabilize into a steady-state response. This initial drop in energy dissipation, known as preconditioning, is a common characteristic of many rubbery (*51*) and biological (*52*) materials. Under cyclic compression with progressively increasing strain (from 10% to 50%), the material continues to follow the primary loading path until extensive damage occurs, as long as the strain in each cycle exceeds the maximum strain of the previous cycle (Fig. S6). These observations mirror the Mullins effect seen in certain rubbers (*51*).

T-6-ring and T-6-HEX exhibit comparatively lower stiffness and can withstand maximum strain without damage to their constituent units (Figs. S6A,B), unlike PAMs composed of polyhedral particles (Figs. S6C,F). This discrepancy can be attributed to the fact that jamming in polyhedron-based PAMs occurred at lower strain values than in 2D polygons or rings (Figs. S6A-F). Additionally, the larger number of kinematic DOFs and the coupled deformation modes of rings and hexagonal particles resulted in lower overall PAM's stiffness. This phenomenon is reminiscent of powders, where sphere-like particles (analogous to PAMs composed of polyhedral particles) are found to exhibit higher compressive strength and elastic modulus, as compared to flake-like particles (analogous to T-6-ring and T-6-HEX) (*53*). Concurrently, it is also possible to create PAMs with polygonal particles exhibiting higher stiffness, akin to PAMs composed of polyhedral particles, by adjusting the kinematic DOFs. For example, J-4-ring and J-4-square structures demonstrated relatively higher stiffness (Figs. S6D,E). Therefore, PAMs can be engineered to achieve greater stiffness by either (i) increasing the number of edges of the particles or (ii) disrupting the kinematic DOFs among particles.

Anisotropy in PAMs and Their Mechanical Behavior

PAMs can inherit intrinsic anisotropies from their crystalline network-based designs. One common manifestation of this anisotropy is the orientation-dependent wetting behavior observed when PAMs are fabricated into spherical shapes. Under gravity, the structural outlines of PAMs will depend on their orientation relative to their crystalline axes (Figs. 2A-C, Movie S5). This anisotropic behavior is markedly different from that of traditional solids, fluids, or granular media, providing PAMs with unique adaptive capabilities.

To better quantify the impact of structural anisotropy on the mechanical behavior of PAMs, we performed a series of simple shear experiments on J-4-ring PAM samples exhibiting varying degrees of structural anisotropy. Specifically, we manipulated the orientation of cubic boundary conditions relative to the periodic lattices by rotating them 15, 30, and 45 degrees (Fig. S20). Owing to the symmetry properties of the J-4-ring PAM, a rotation of 15 degrees corresponds to an effective rotation of 75 degrees, while a 30-degree rotation corresponds to 60 degrees.

During the simple shear experiment, the plane of structural rotation was designated as the plane for y-direction shear, with the orthogonal plane corresponding to x-direction shear. At zero degrees of rotation, the structure exhibited symmetry about the axis perpendicular to the shearing direction on the shear plane, resulting in equal displacements in the positive and negative y-directions at near-zero shear force. As the rotation increased to 15 and 30 degrees, this symmetry was disrupted, leading to more pronounced anisotropic behavior and variations in the displacements observed. At 45 degrees of rotation, the inherent orthotropic symmetry of the structure's topology reinstated the symmetry, producing displacements in both directions similar to those at zero degrees, but with different displacement values, all corresponding to near-zero shear force. Since the structural rotation was confined to the y-direction shear plane, the symmetry on the x-direction shear plane remained unaltered. Consequently, the structure exhibited identical displacements in the positive and negative x-directions at near-zero shear force, independent of the applied rotation. These displacement values showed only minimal variation with increasing rotation, underscoring the invariable response of the x-direction shear despite the structural changes occurring in the y-shear plane.

Randomness and Anisotropic of PAMs' Mechanical Behavior

One distinguishing characteristic of PAMs is their remarkable ability to adapt to environmental conditions, including external forces. This adaptability makes their global shapes and mechanical behaviors highly sensitive to their initial configurations. This effect typically leads to random mechanical responses to identical loading conditions.

Due to the infinite possible ways PAMs can adapt their particle arrangements, their mechanical behaviors, such as responses to compressive loads, are highly sensitive to the nuances of their initial configurations. For example, a cubic J-4-ring PAM sample subjected to the same compressive loading condition ten times, but with slightly different initial configurations each time, can exhibit varying moduli despite similar loading-unloading profiles (Fig. S8). Notably, this limited degree of randomness arises from the complex interplay between initial particle arrangements and mechanical responses.

Rheology Test

Rheology is the study of how materials flow and deform under applied forces, focusing particularly on the behavior of complex fluids and soft solids. This discipline examines the responses of materials to stress, strain, and time, thereby revealing intricate dynamics related to flow and deformation. Such insights are pivotal in understanding material properties such as viscosity, elasticity, and plasticity. In this context, we define several key rheological terms relevant to our study:

*Complex modulus ($G^*$)* – In oscillatory tests, $G^*$ is defined as the ratio of the applied (or measured) stress amplitude to the measured (or applied) strain amplitude. $G^*$ serves as a precise quantitative indicator of a material's ability to resist deformation.

*Phase angle ($\delta$)* – Phase angle is a measure of the balance between viscous and elastic behaviors in a material, ranging from 0° (indicative of a purely elastic material) to 90° (characteristic of a fully viscous material).

*Storage modulus ($G'$)* – Defined by the equation $G' = G^*\cos(\delta)$, this modulus quantifies the elastic component of a material's viscoelastic behavior, indicating how much energy is stored and recovered in one cycle of deformation.

*Loss modulus ($G''$)* – This modulus is defined by the formula, $G'' = G^*\sin(\delta)$, this modulus reflects the viscous component of a material's viscoelastic behavior, representing the energy dissipated as heat within the material during deformation.

*Complex viscosity ($\eta^*$)* – This parameter evaluates the overall resistance a material presents to flow under oscillatory conditions. It is dependent on the angular frequency ($\omega$) and is calculated using the equation $\eta^* = G^*/\omega$. Complex viscosity effectively encapsulates both the viscous and elastic responses of the material, providing a comprehensive measure of how the material behaves under dynamic stresses.

In the case of this study, the storage modulus (G') indicates the energy stored as the deformation of PAMs, while the loss modulus (G") reflects the energy dissipated through internal friction within PAMs during the test. A material exhibits characteristics of a viscoelastic solid when $G' > G''$, indicating predominant elastic behavior. Conversely, it behaves as a viscoelastic fluid $G' < G''$, where viscous behavior is more pronounced. Our tests confirm that PAMs behave as viscoelastic solids, as evidenced by consistently higher G' than G" in all oscillatory tests, including oscillatory amplitude and frequency sweeps.

All rheology tests of PAMs showed varying G' and G", as well as $\eta^*$, exhibiting a transition from shear-thinning to shear-thickening fluid behavior. During the amplitude and frequency sweeps, both the moduli and viscosity initially decrease and then increase as the shear rate increased (Fig. 3). This distinctive response is attributed to the semi-granular/liquid-like nature of PAMs led to complex interactions between constituent particles, resulting in this unique rheological behavior.

In contrast, when the same amplitude sweep tests were performed on a simple cubic truss lattice with the same material, manufacturing technique, and outlining dimensions as PAMs, both moduli and complex viscosity remain constant regardless of increase in shear rate (Fig. S16), demonstrating a typical behavior of a solid.

**Supplementary References**

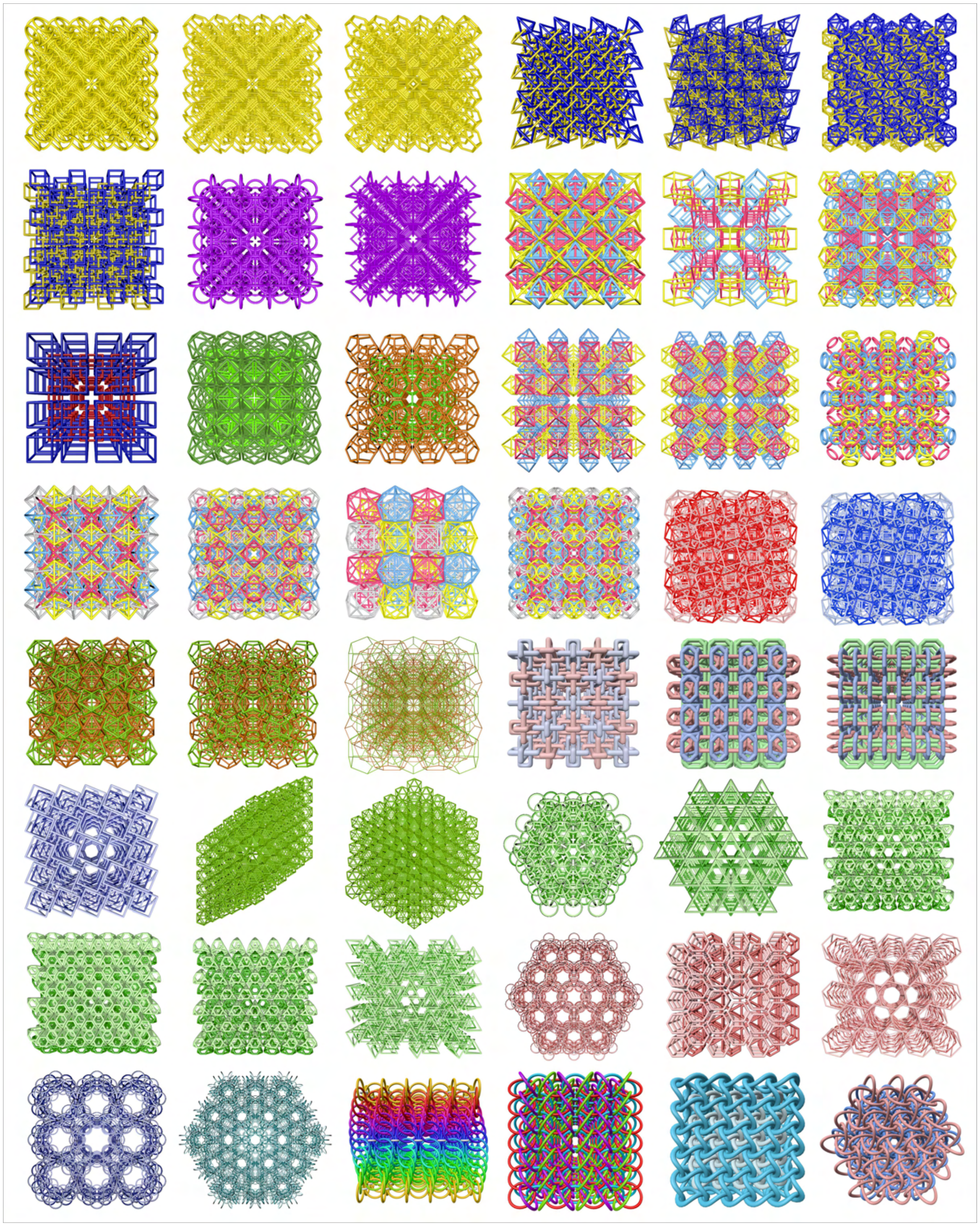

**Fig. S1.**

**Selected library of PAMs composed of single type of particles.** Different colors highlight local structural features that may be visually challenging to distinguish, including particle geometries, orientations, and linking types. PAMs consist of more than one types of particles are not listed in this library.

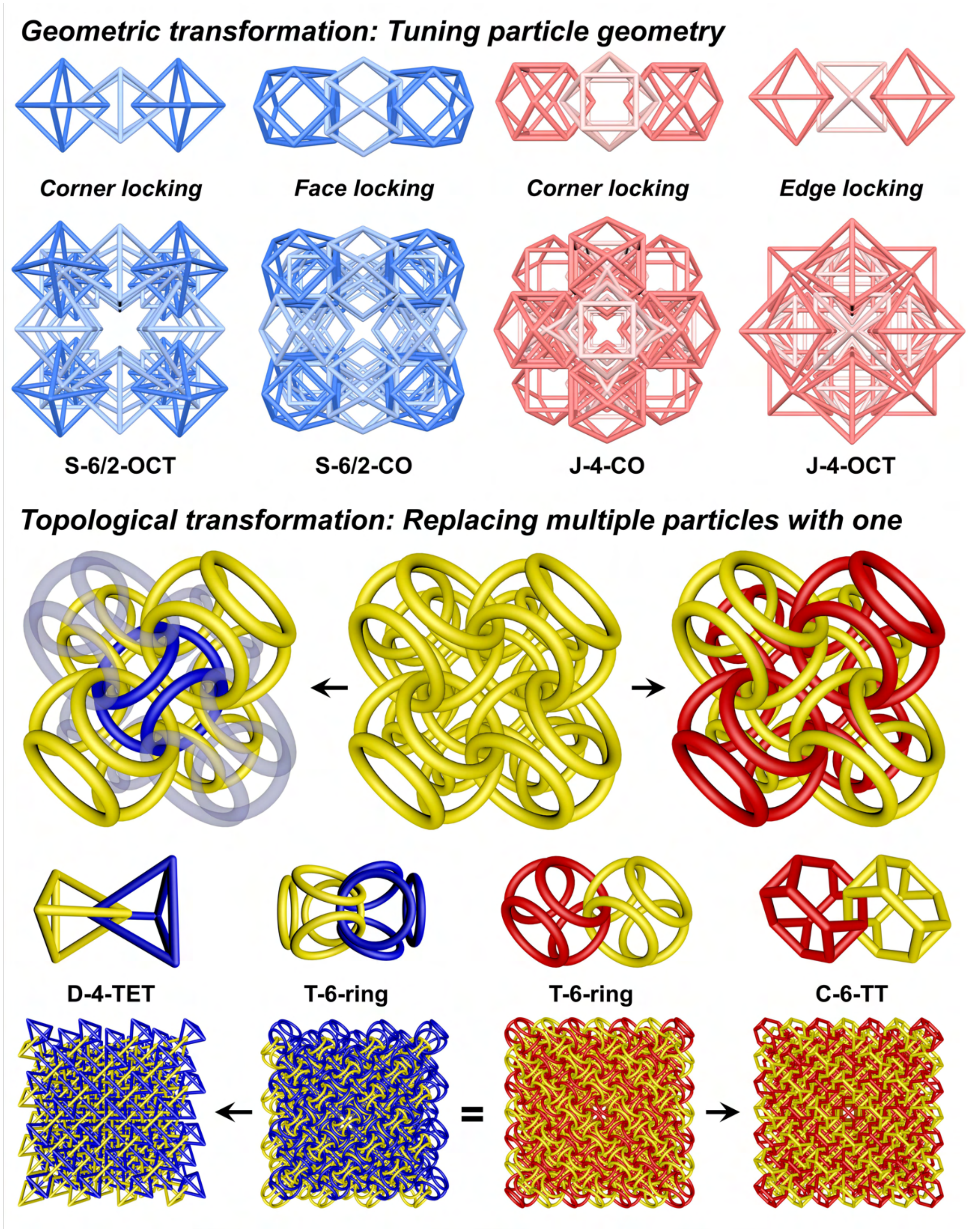

**Fig. S2.**

**Structural tunabilities of PAMs.** (I) Tuning particle geometry of PAMs with the same topology. PAMs composed of octahedral (OCT) and cuboctahedral (CO) particles are linked in S-6/2 and J-4 topologies. By truncating octahedra into cuboctahedra, existing local contact mechanisms of PAMs can be drastically altered, namely transitioning from corner-to-corner locking into face-to-face locking or edge-to-edge locking. (II) Replacing multiple particles with one (geometric transformations among T-6-ring, D-4-TET, and C-6-TT). On the left side, T-6-ring is divided into two sets of tetrahedral clusters (yellow and blue), arranged in a *dia* topology. Replacing each cluster with one tetrahedral particle results in a D-4-TET. On the right side, T-6-ring is divided into two sets of tetrahedral clusters (yellow and red), clusters arranged in a *pcu* topology. Replacing each cluster with one truncated tetrahedral particle results in a C-6-TT.

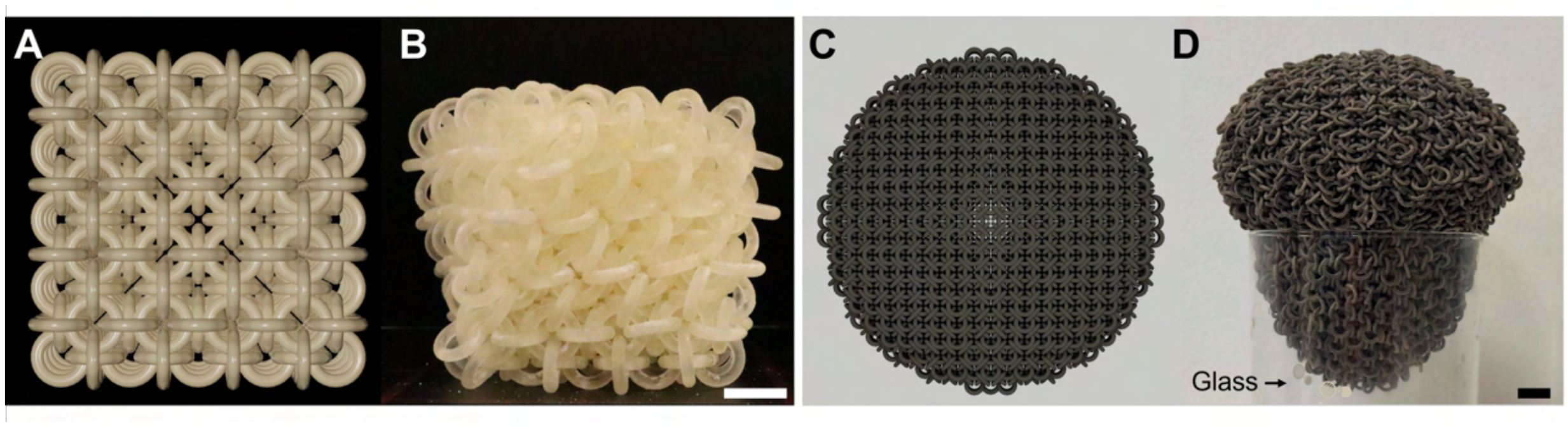

**Fig. S3.**

**Relaxation behavior of PAMs under gravity.** The global outlines of J-4-ring PAMs are designed as a cube composed of 540 ring particles (A,B) and a sphere composed of 9414 ring particles (C,D). Scale bars: 1 cm.

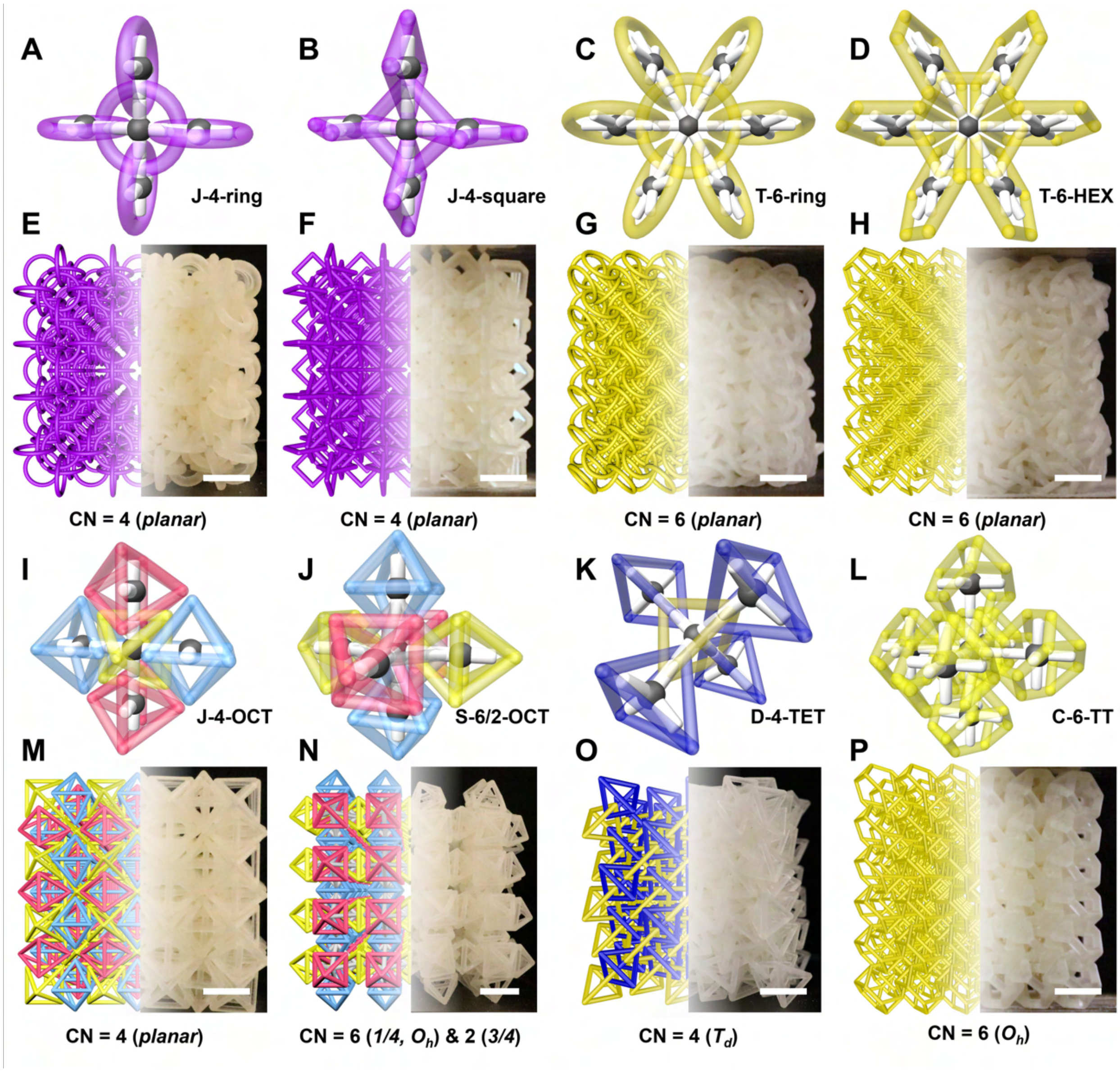

**Fig. S4.**

**Catenation configurations and fabrication of PAMs.** (A-D) Schematic representations of J-4-ring, J-4-square, T-6-ring, and T-6-HEX PAMs' local configurations, where J-4-ring and J-4-sqr have planar CN of 4, T-6-ring and T-6-HEX have planar CN of 6. (E-H) Illustration and corresponding photos of expanded J-4-ring, J-4-SQR, T-6-ring, and T-6-HEX PAM samples. (I-L) Schematic representations of J-4-OCT, S-6/2-OCT, D-4-TET, and C-6-TT PAMs' local configurations, where J-4-OCT has planar CN of 4, S-6/2-OCT has octahedral ($O_h$) CN of 6 and 2, D-4-TET has tetrahedral ($T_d$) CN of 4, C-6-TT has $O_h$ CN of 6. (M-P) Illustration and corresponding photos of expanded J-4-OCT and S-6/2-OCT, D-4-TET, and C-6-TT PAM samples. Scale bars: 1 cm.

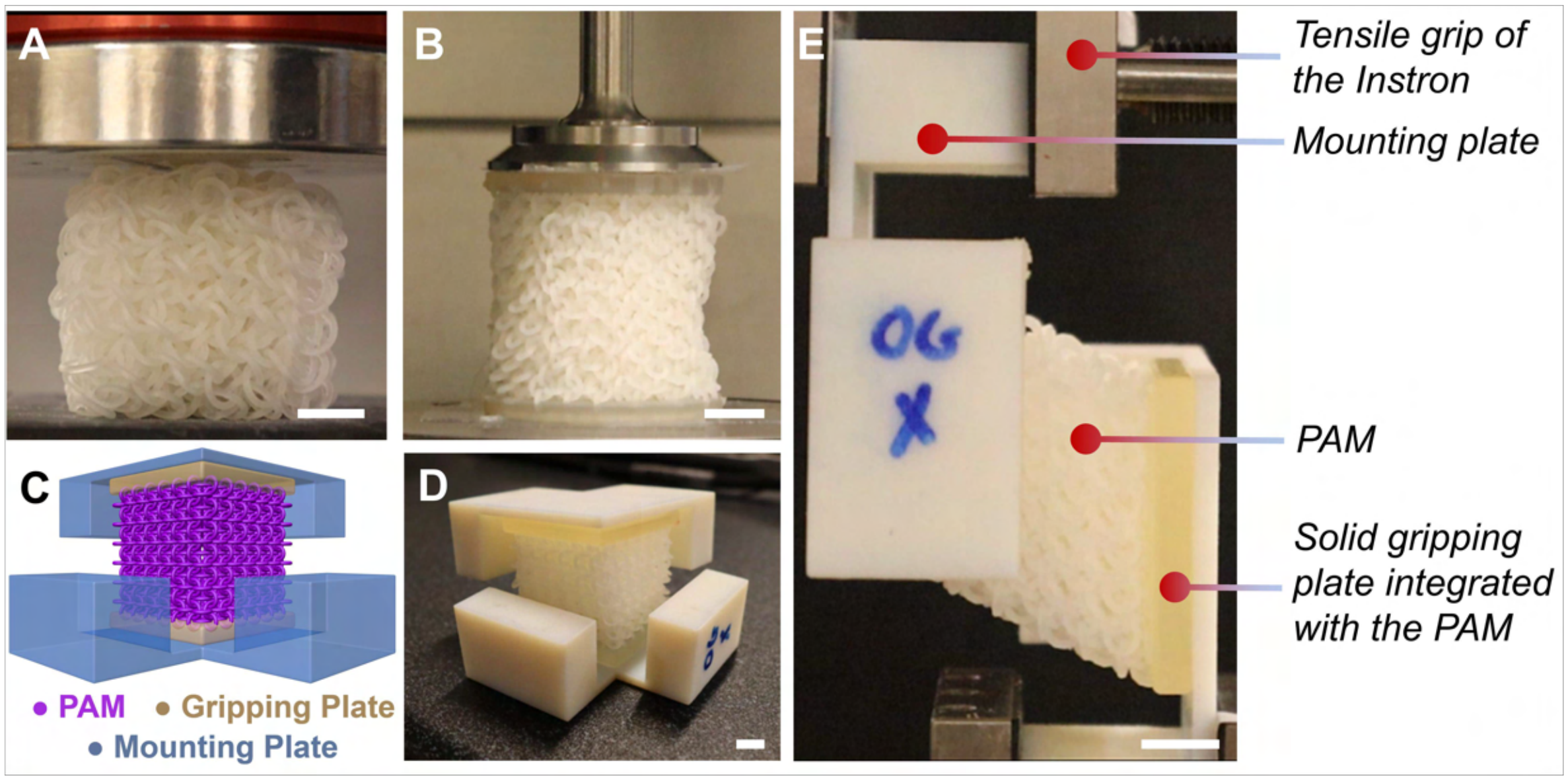

**Fig. S5.**

**Experimental setups for mechanical characterizations.** (a) Uniaxial compression test setup, (b) Rheology test setup, (c) Design and positioning of the mounting plate on the PAM specimen to secure the shear test sample in the Instron testing machine, (d) Assembly of the mounting plate with the solid gripping plate integrated into the shear test specimen, and (e) Shear test specimen positioned in the Instron machine utilizing the solid gripping plate and mounting plate. Scale bars: 1 cm.

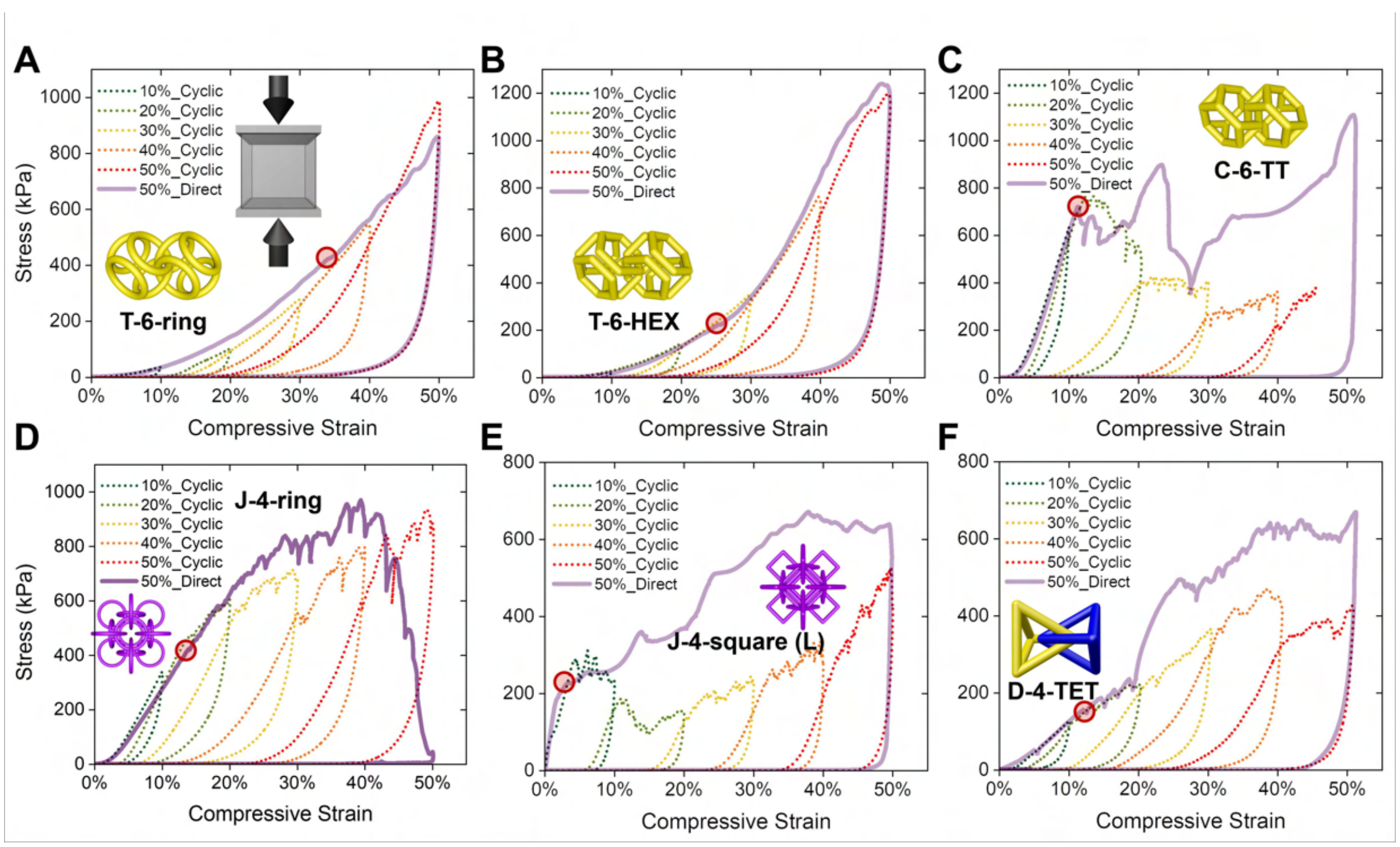

## Fig. S6.

**Uniaxial compression test of PAMs.** (A-F) Stress-strain results of six PAMs from cyclic loading (10%, 20%, 30%, 40%, and 50%) and direct loading (50%): (A) T-6-ring, (B) T-6-HEX, (C) C-6-TT, (D) J-4-ring, (E) J-4-square, and (F) D-4-TET. Red circles indicate the onset of significant fatigue, calculated from the first derivative tests.

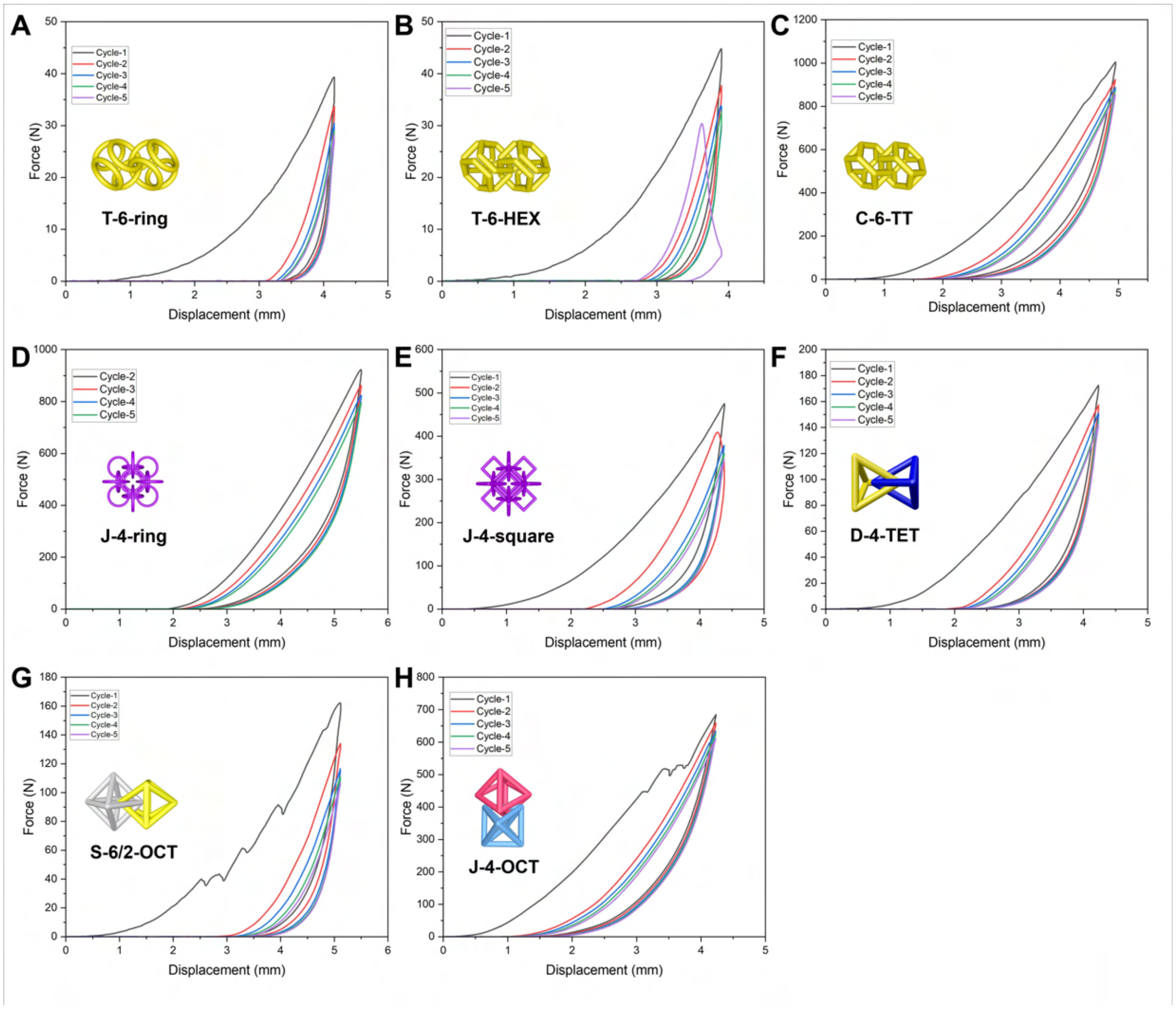

## Fig. S7.

**Cyclic compression of all PAMs at same loading conditions (10% strain).** A series of force-displacement response from cyclic compression tests on all discussed PAMs in this study at 10% strains illustrating the preconditioning of the PAMs during the initial loading cycles.

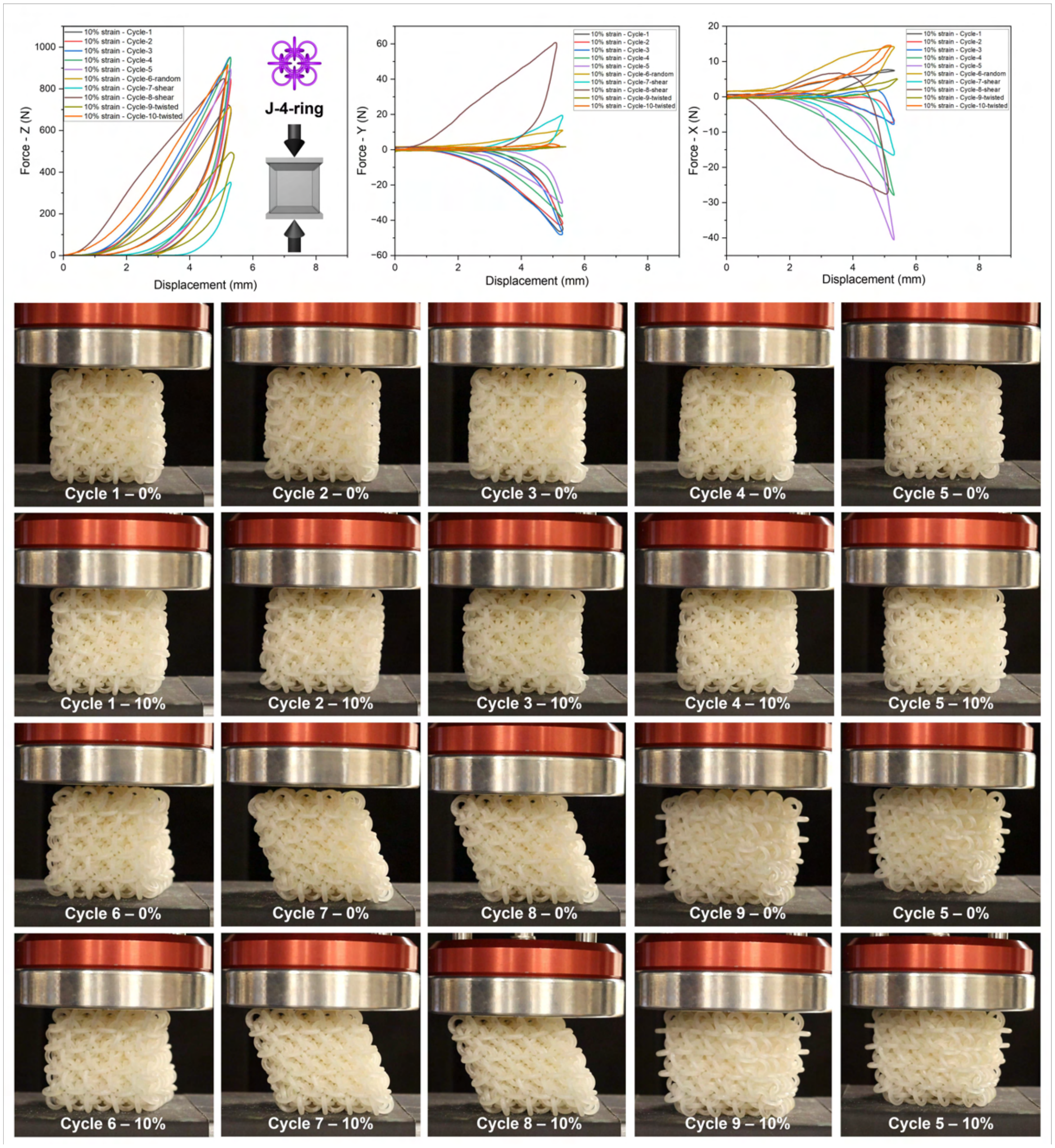

**Fig. S8.**

**Force-displacement response of the J-4-ring PAM under uniaxial compressive loading across various initial configurations.** The force responses were measured using a triaxial force sensor, which recorded forces along the loading direction (Z-axis) and two perpendicular directions (X and Y axes). The panel of images displays the different initial configurations tested, including upright, random, sheared, and twisted arrangements. The plots clearly illustrate the influence of each configuration on the global force-displacement behavior along the Z-axis. Additionally, the forces recorded in-plane (X and Y directions) display a high degree of randomness.

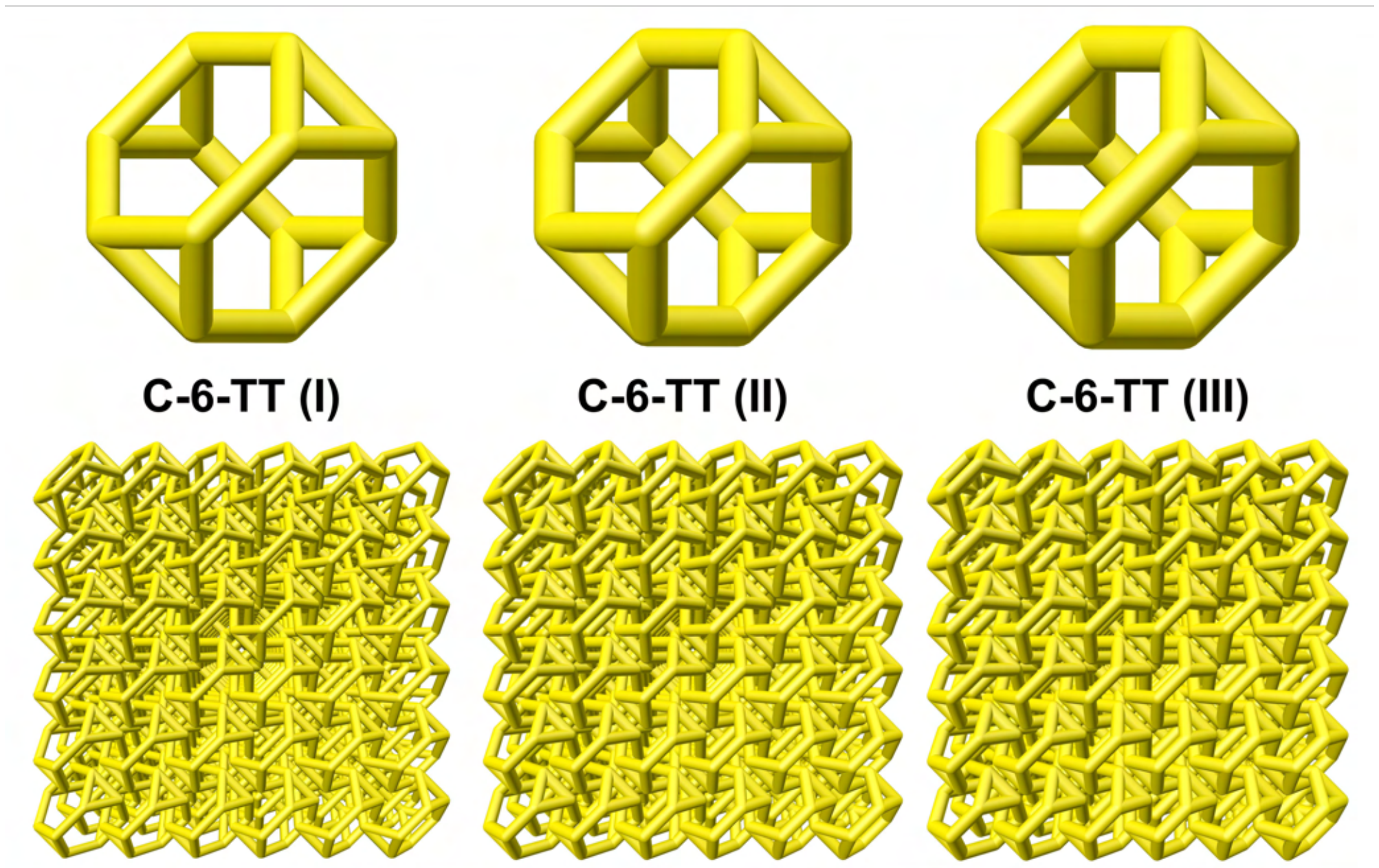

**Fig. S9.**

**Three C-6-TT PAMs with different volume fractions.** Three PAM structures are designed with the same particle size (truncated tetrahedra), only different in their beam thicknesses, with a ratio of 7:8:9. These PAM structures are fabricated at both macro- and microscale, in order to compare the scalability of PAMs' mechanical responses.

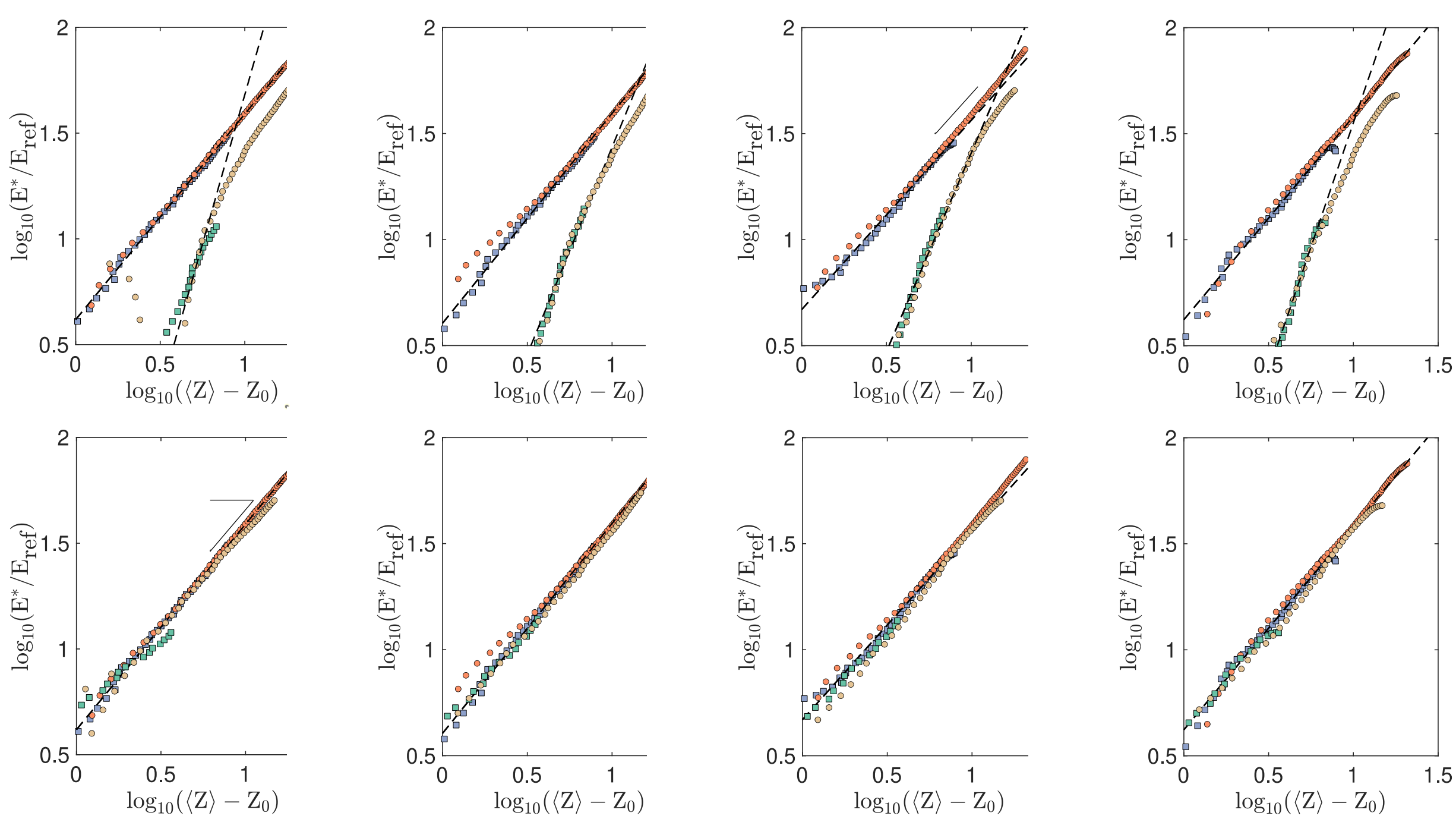

**Fig. S10.**

**Results of E* using different polynomial fit to simulated stress-strain responses of T-6-ring sample. a, c, e,** and **g** Using polynomial of second, third, fourth and fifth degree, respectively, but all with $Z_0 = 2$ for the two "As is" cases. **b, d, f,** and **h** Using polynomial of second, third, fourth and fifth degree, respectively, but all with $Z_0 = 5$ instead for the two "As is" cases. $Z_{\mathrm{ref}}$ is a constant selected as 1 kPa, used for the sole purpose of normalizing **E*** to be dimensionless.

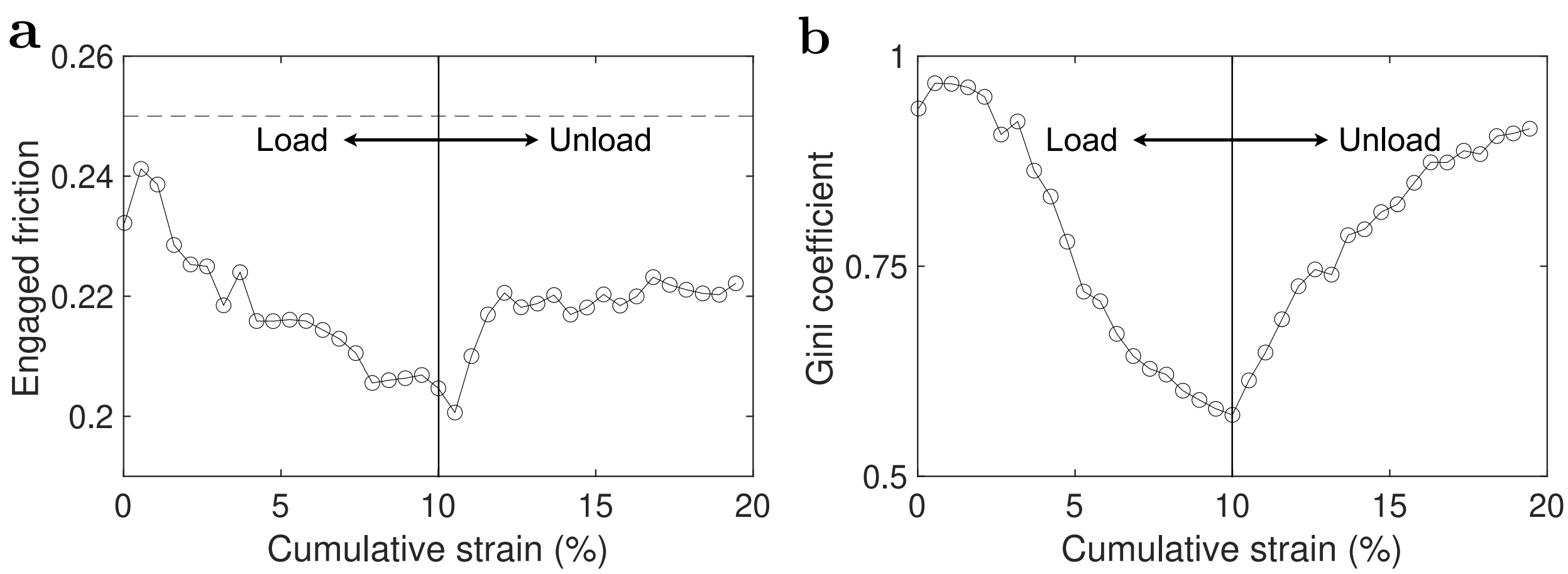

**Fig. S11.**

**Additional particle-scale contact analysis from LS-DEM simulation of T-6-ring under uniaxial compression. a** Variation of mean "engaged" friction between particles as a function of cumulative strain. **b** Variation of the Gini coefficient of contact forces as a function of cumulative strain.

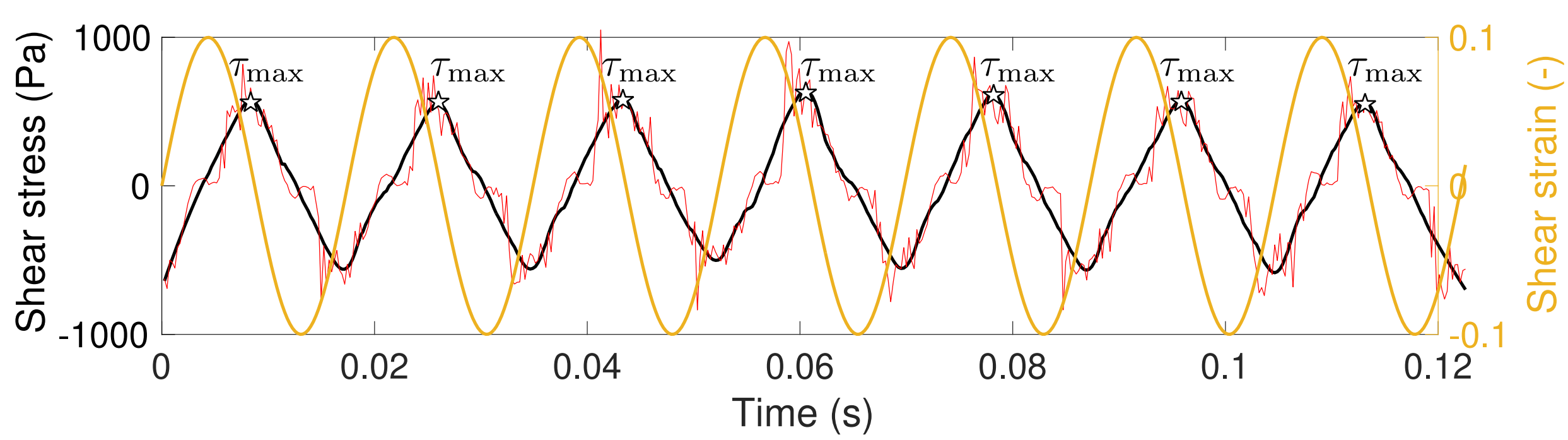

**Fig. S12.**

**A representative rheology simulation output of J-4-ring under an excitation frequency of 360 rad/s.** Left axis shows the variation of the shear stress (raw output in red line while smoothed data in black solid line) as a function of time calculated at the top boundary. Right axis shows the imposed sinusoidal shear strain with an amplitude of 0.1.

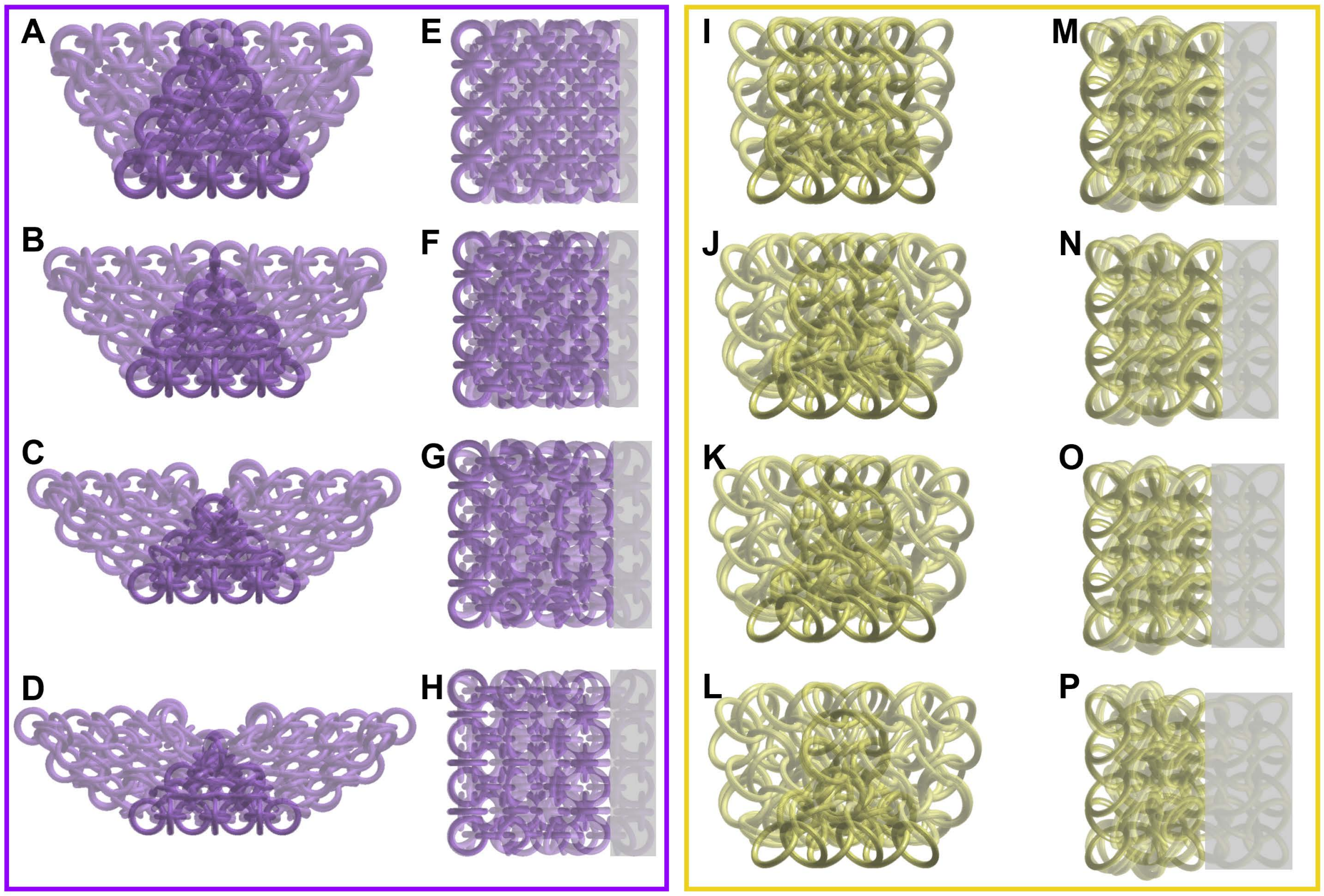

**Fig. S13.**
**Simulated critical jamming strains of PAMS.** (A-P) Simulation results displaying J-4-ring (A-H) and T-6-ring (I-P) PAMs at their corresponding critical jamming strains.

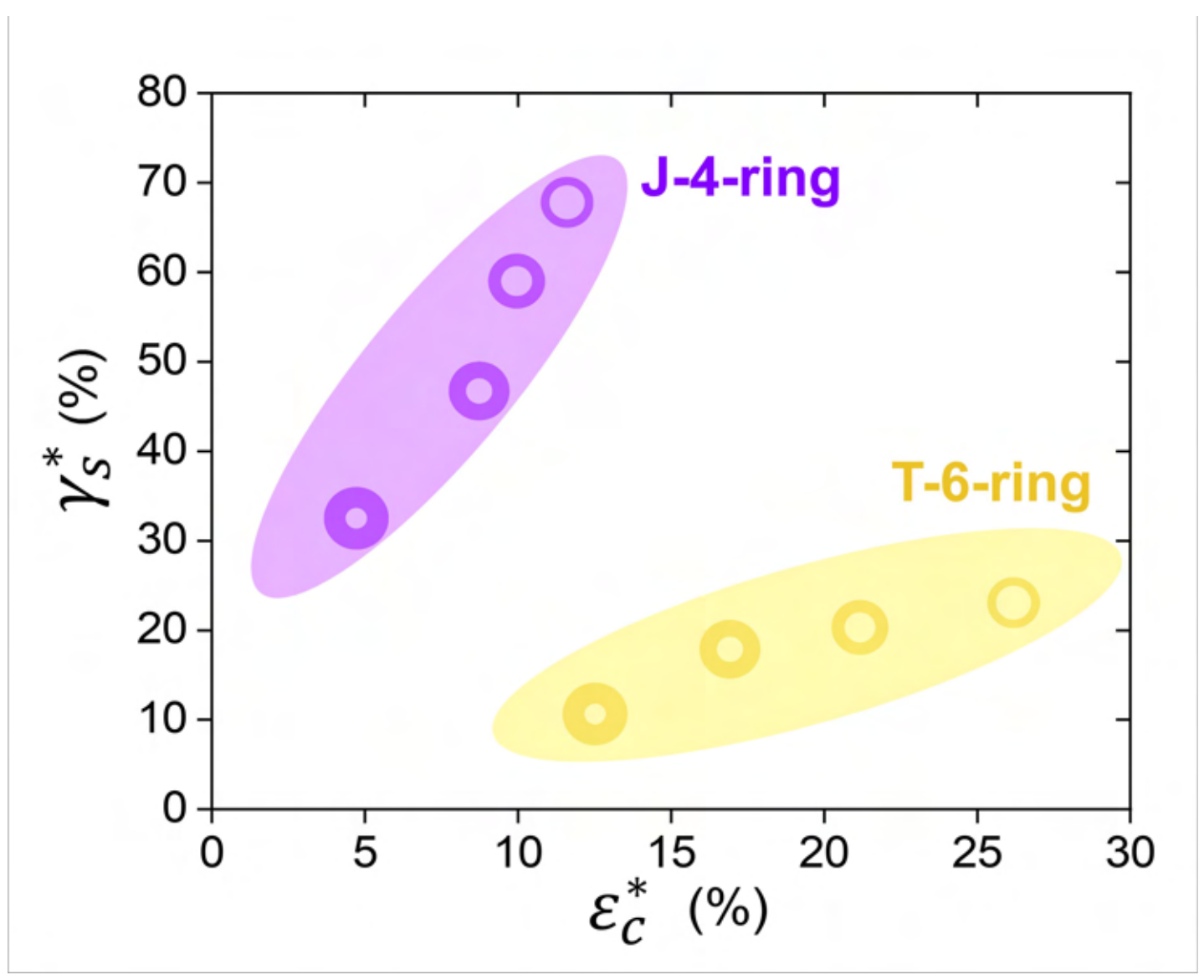

**Fig. S14.**

**Simulated critical jamming strain of PAMs.** The simulated critical jamming strain under both shear and compressive loading conditions is analyzed as a function of ring thickness and topology, as described in Fig. 4. The trends and values match the experimental results shown in Fig. 4B.

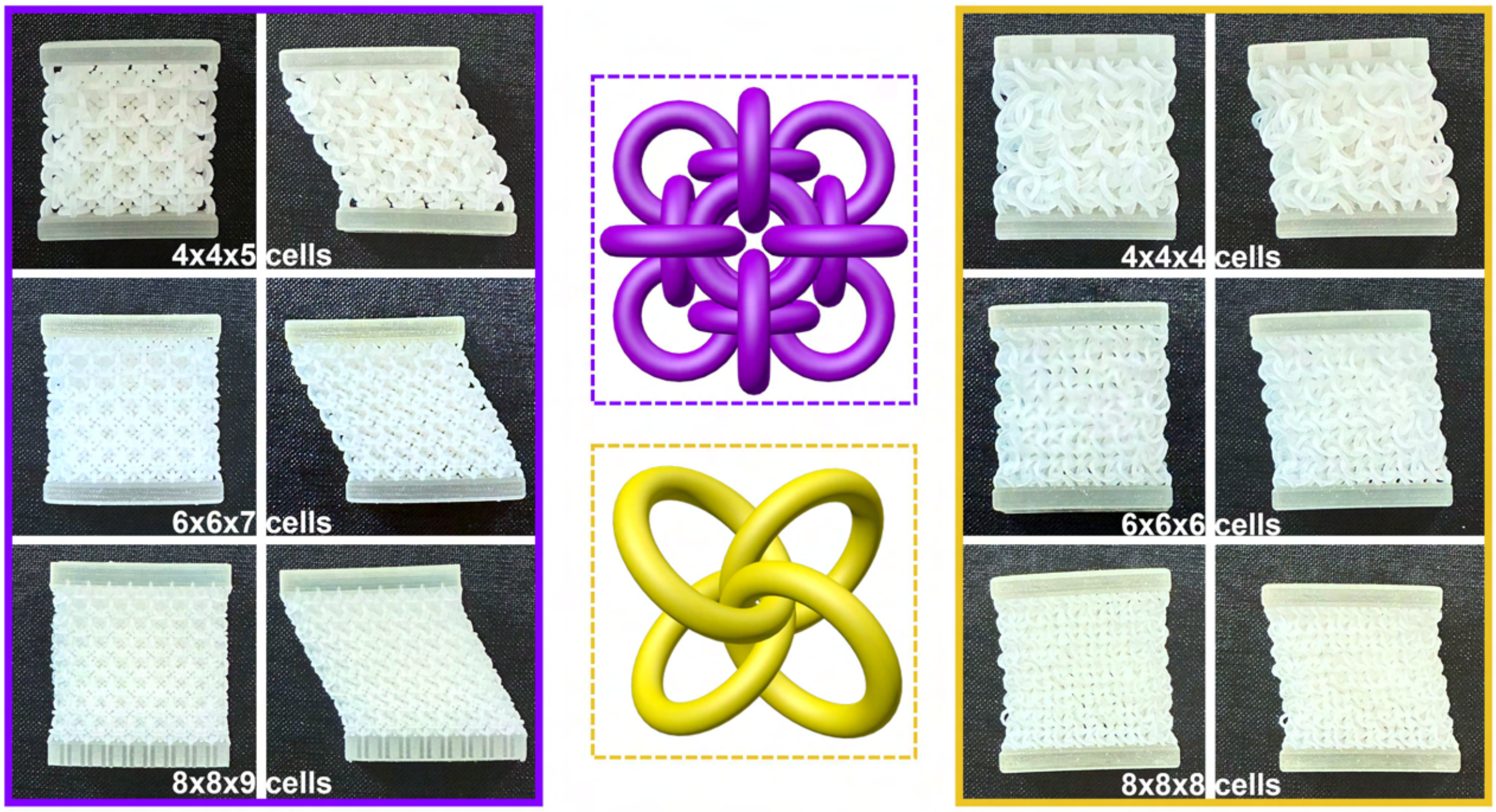

**Fig. S15.**

**Maximal critical shear jamming strains of J-4-ring and T-6-ring PAMs.** Top-down photographs show J-4-ring (left) and T-6-ring (right) PAMs, fixed with top and bottom gripping plates, horizontally placed on a flat surface. From top row to bottom row, PAMs are designed within the same domain, but with increasing numbers of smaller unit cells. Left columns show the natural state of the PAMs, while right columns show PAMs at their critical shear jamming strains.

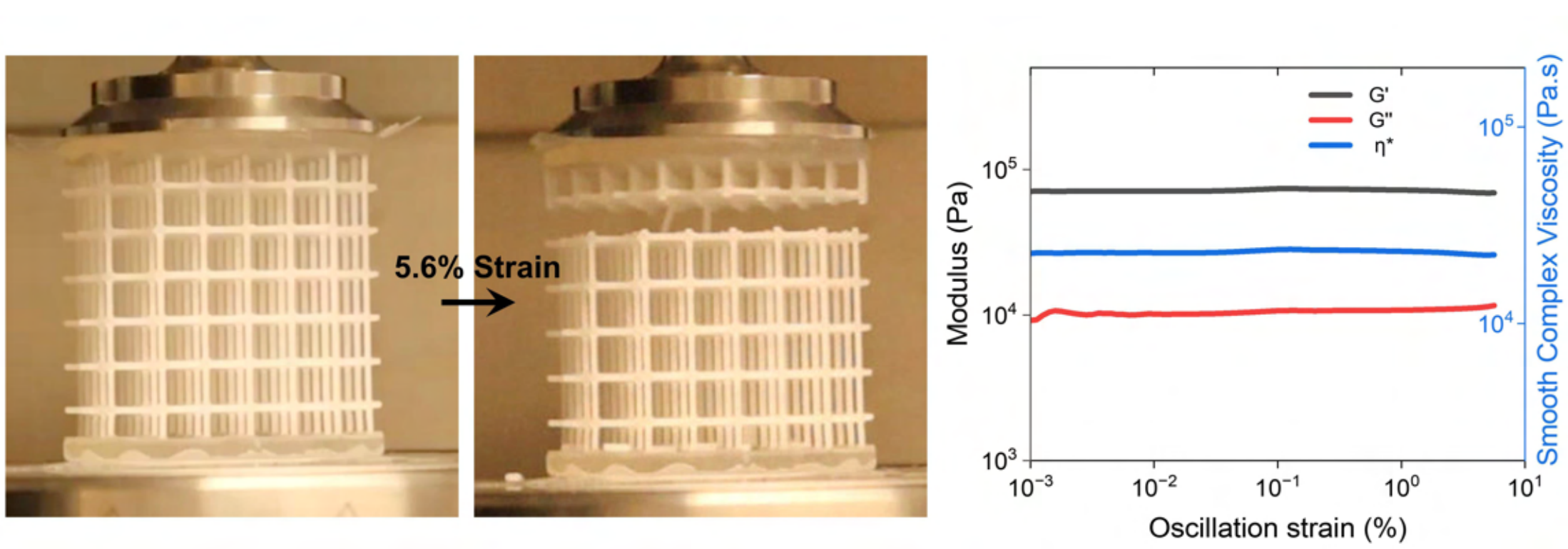

**Fig. S16.**

**Rheology test of simple cubic truss lattice.** Experimental setup and results of an amplitude sweep test using a simple cubic truss lattice, with consistent configurations as the tests on PAMs. The test was automatically stopped at an oscillation strain of 5.6%, due to a catastrophic failure developed across one layer in the sample.

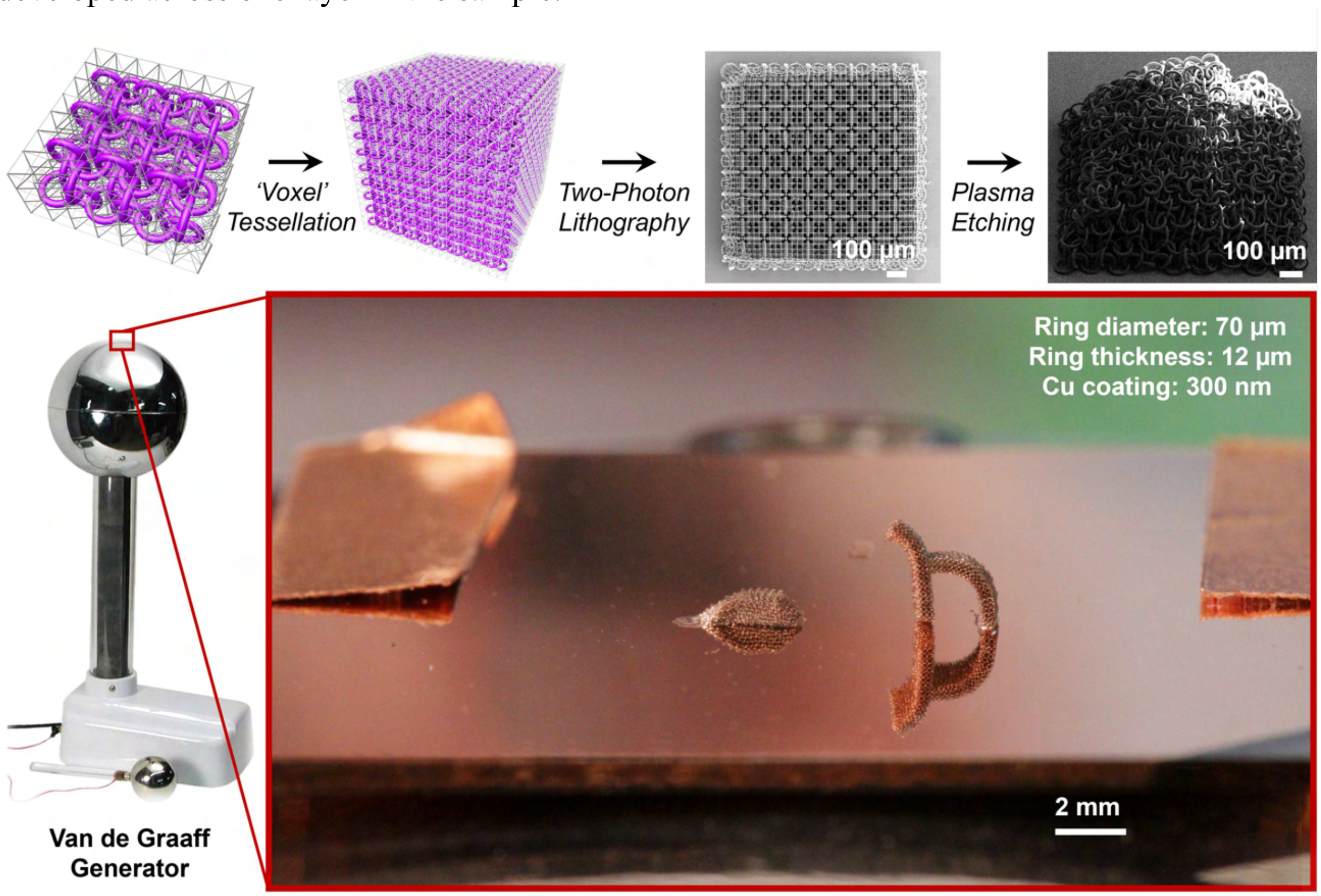

**Fig. S17.**

**Fabrication and electrostatic reconfiguration of μ-PAM.** Top row shows the design, two-photon lithography, and plasma etching process of a J-4-ring PAM. Bottom panel illustrates Cu-plated μ-PAM samples attached to ITO-coated glass substrates, placed atop a Van de Graaff generator.

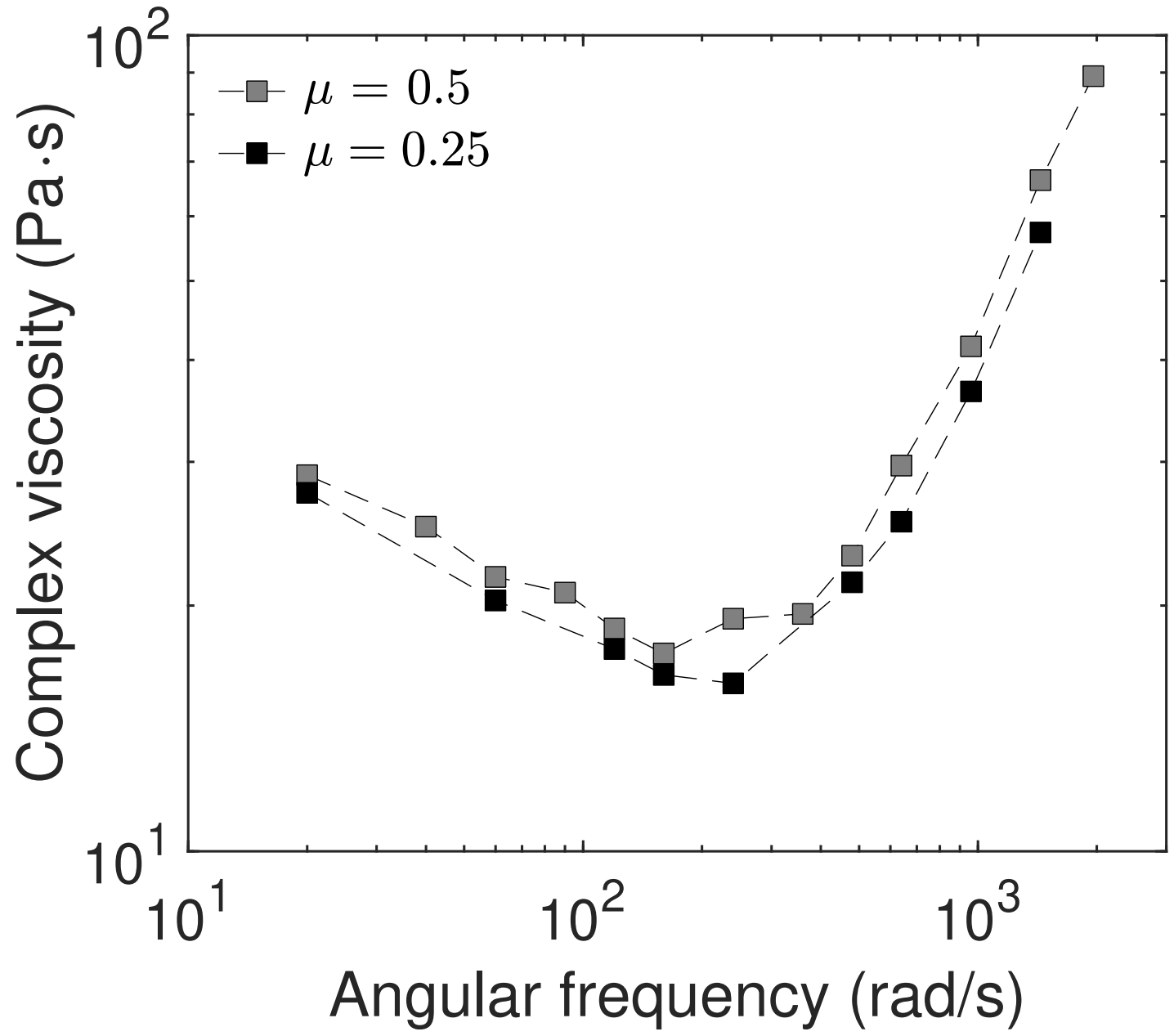

**Fig. S18.**
**Effect of different friction coefficient in the LS-DEM simulation of PAM's rheological behavior.**

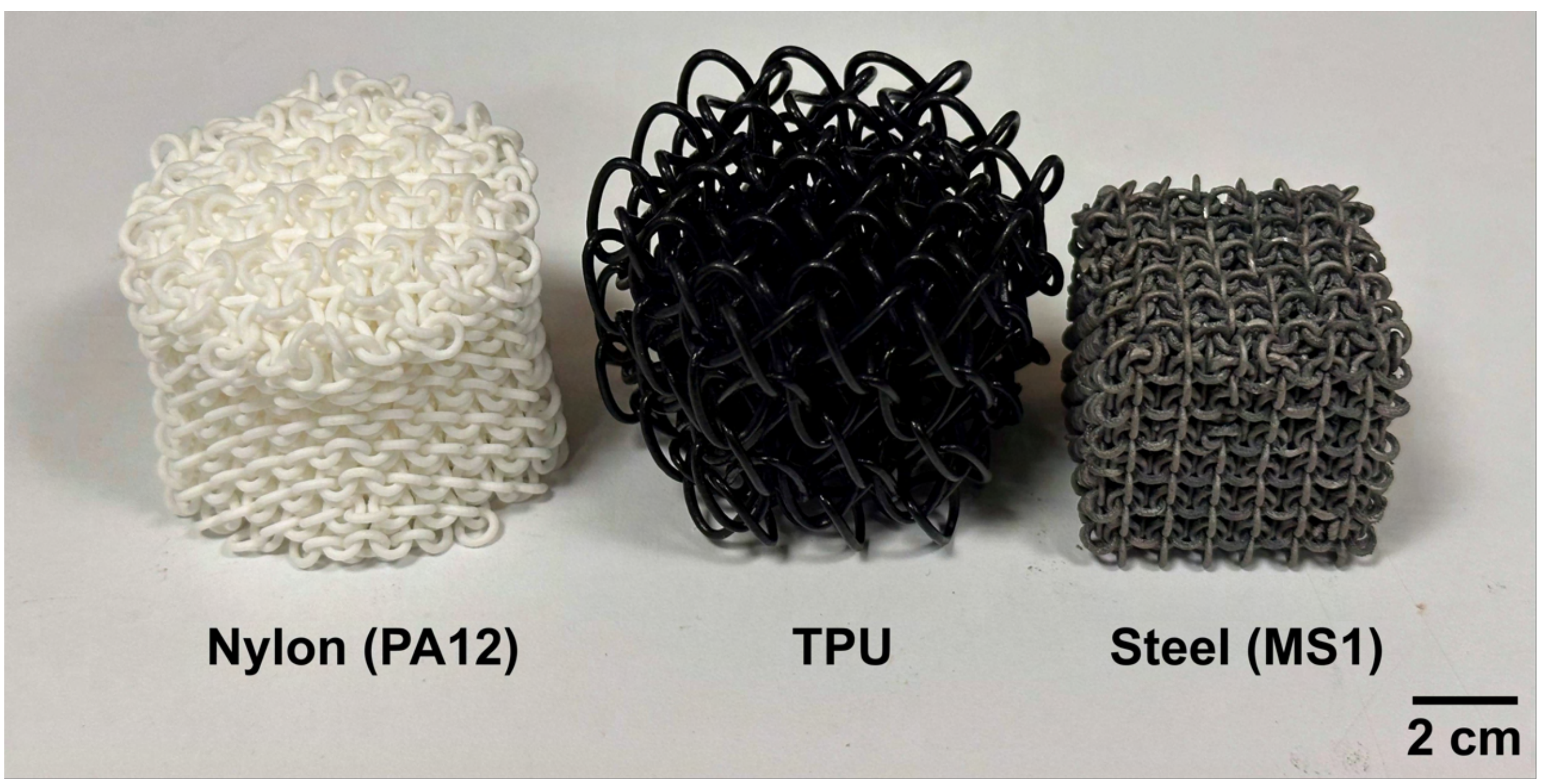

**Fig. S19.**
**PAMs fabricated using varied material compositions.** PAMs printed with three different materials and designs using powder laser sintering, a method ideal for support-free structures. The design and material variations result in different macroscopic properties, showcasing the versatility of PAMs.

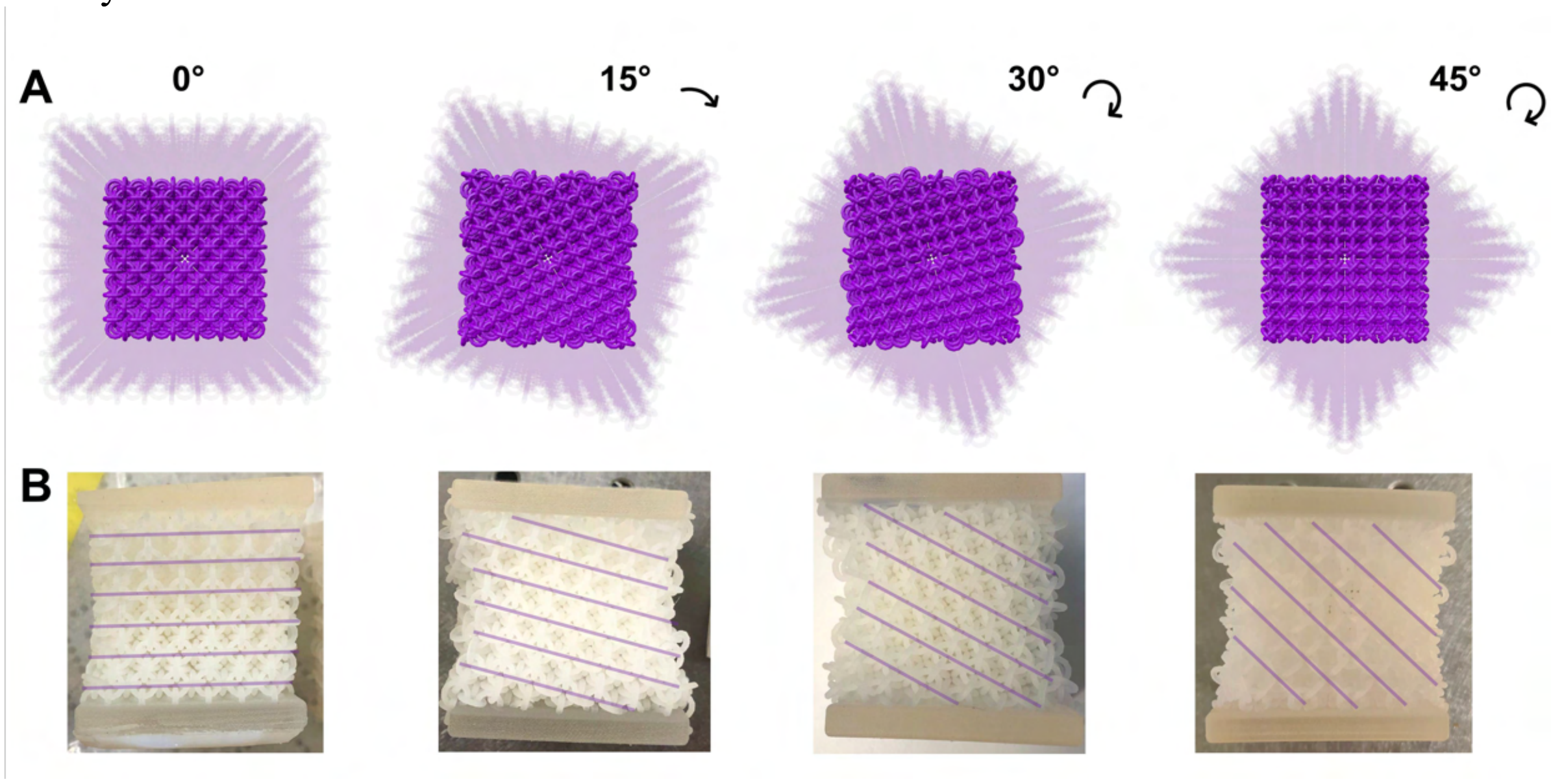

**Fig. S20.**

**Configuration of J-4-ring PAM Samples Under Structural Rotation**. (A) Illustrations showing the cubic boundary conditions of a J-4-ring PAM matrix being rotated at 0°, 15°, 30°, and 45°. (B) Photographs of samples corresponding to each rotation angle depicted in (A).

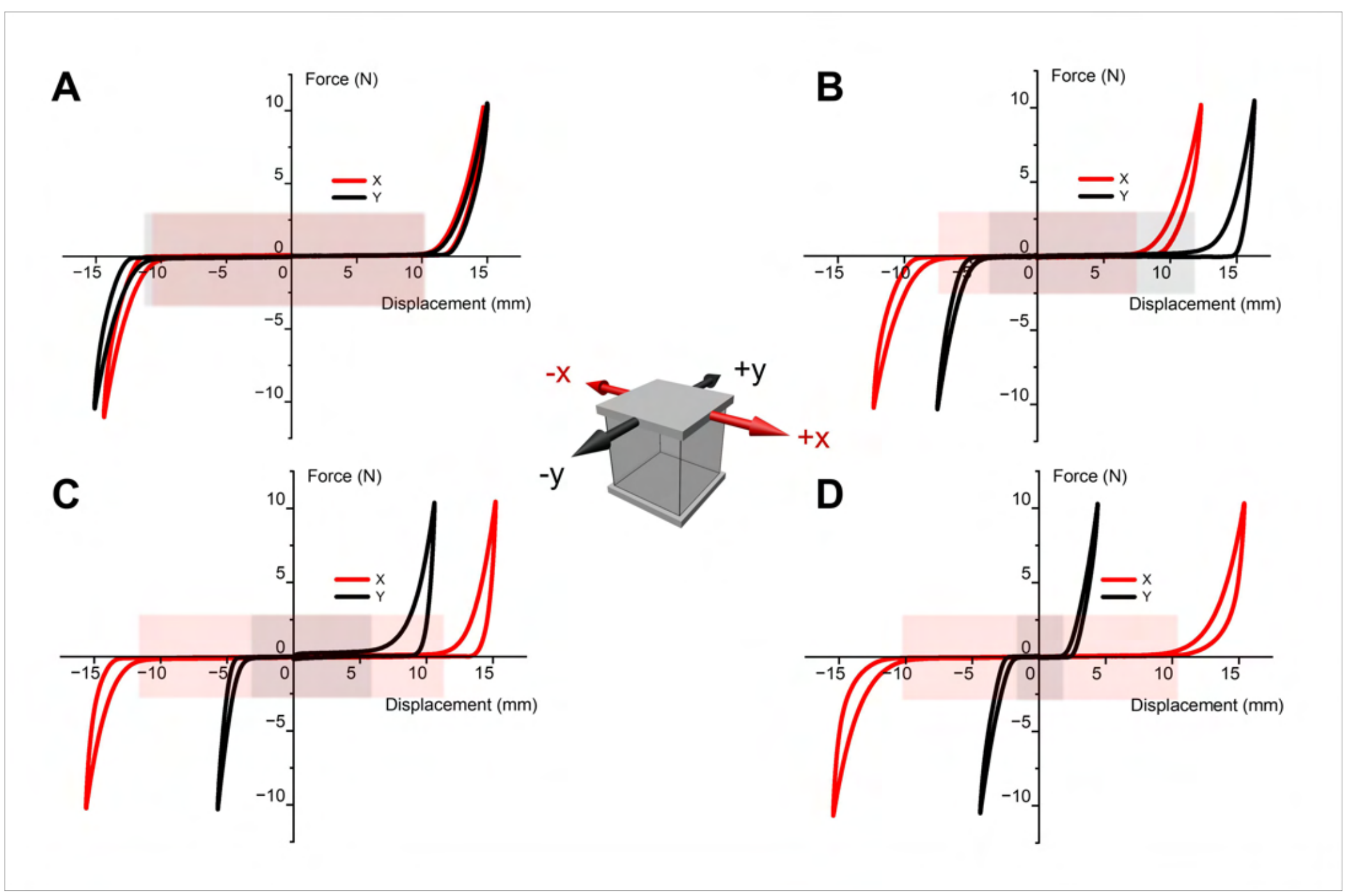

**Fig. S21.**

**Force-Displacement Curves for Rotation Series of J-4-ring PAM Under Varied Shear Conditions.** Shaded regions highlight the range of displacement where shearing force is near-zero.

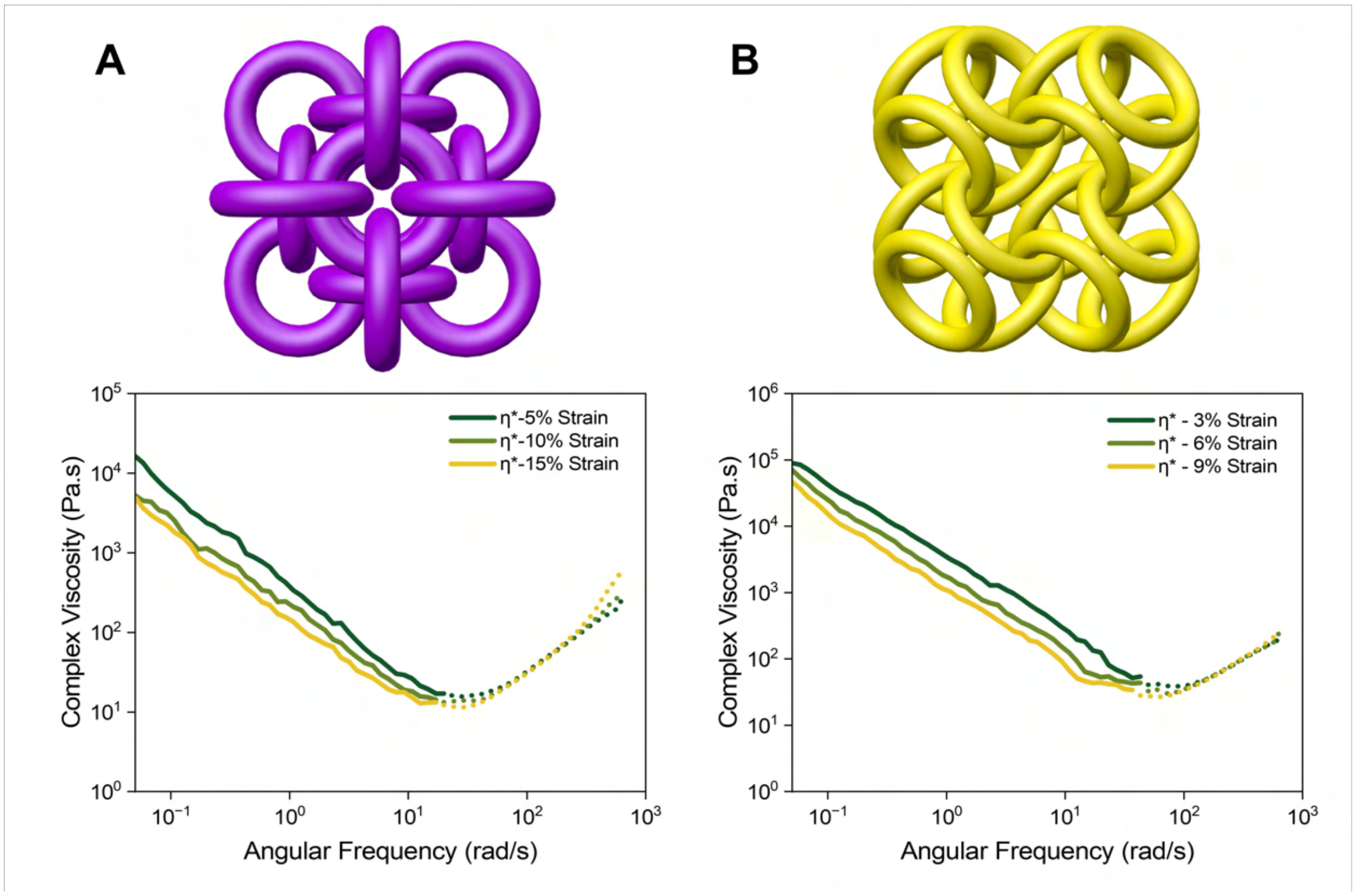

**Fig. S22.**

**Rheological Behavior of J-4-ring (A) and T-6-ring (B) PAMs Under Varying Strain Levels.**

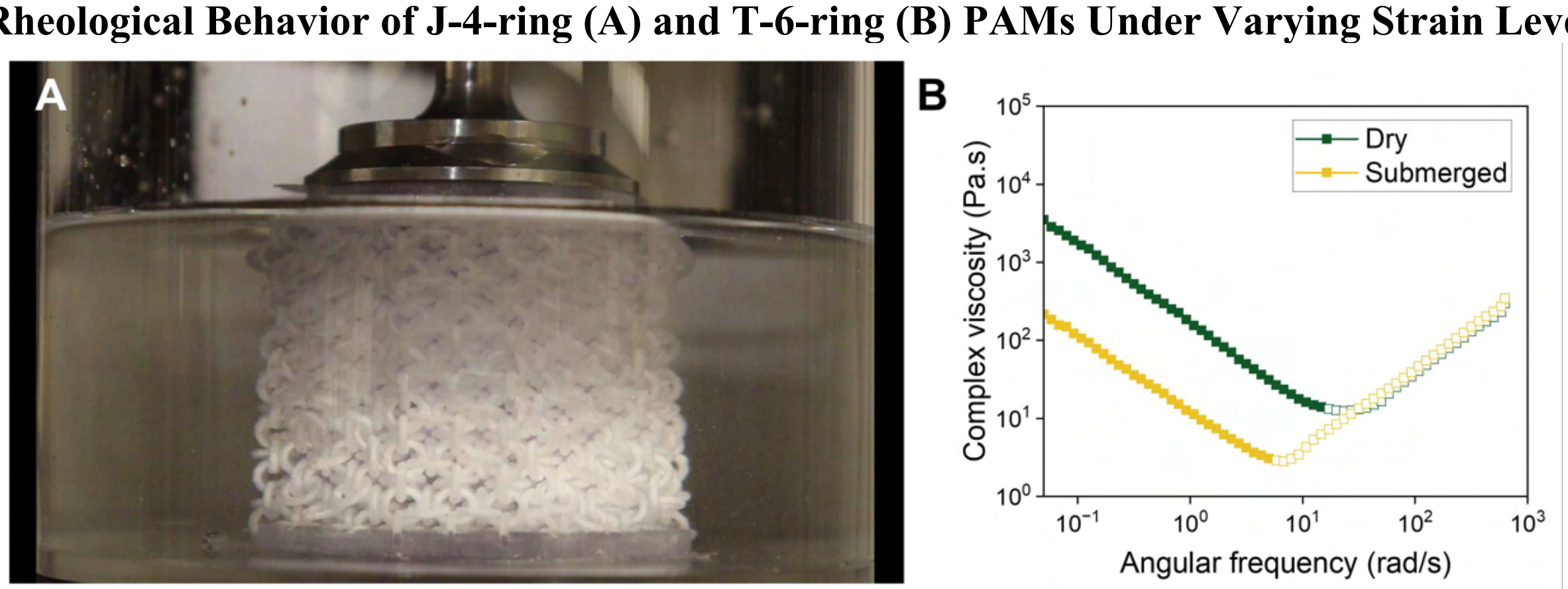

**Fig. S23.**

**Comparison of Complex Viscosity of J-4-ings at Dry State vs Fully Submerged in Water.**
(A) A photograph showing the experimental setup for submerged rheology experiments. (B) Comparison of complex viscosities of dry and submerged J-4-ring PAMs.

**Table S1.**

**Designed and relaxed dimensions of PAMs.**

| Name | Array size ($N \times N \times N$) | Measured width (mm) | Measured depth (mm) | Measured height (mm) | Designed edge length (mm) | Relaxed volume fraction (%) |
|---|---|---|---|---|---|---|
| J-4-ring | $N = 5$ | 54.50 | 54.50 | 54.00 | 55.00 | 31.84 |
| J-4-square* | $N = 5$ | 56.25/42.78 | 56.25/43.79 | 56.25/42.27 | 56.50 | 15.67/35.23 |
| T-6-ring | $N = 6$ | 46.40 | 46.40 | 42.03 | 50.00 | 25.67 |
| T-6-HEX | $N = 6$ | 47.49 | 47.48 | 46.45 | 50.00 | 22.11 |
| J-4-OCT | $N = 4$ | 55.80 | 55.80 | 56.93 | 58.00 | 16.77 |
| S-6/2-OCT | $N = 4$ | 54.20 | 54.20 | 52.40 | 56.00 | 10.87 |
| D-4-TET | $N = 5$ | 59.20 | 59.20 | 58.20 | 60.00 | 12.68 |
| C-6-TT | $N = 6$ | 47.96 | 47.96 | 47.96 | 49.00 | 32.74 |

*J-4-square has two stable configurations (see Movies S7, S8), therefore we listed measured dimensions for configurations L/S.

**Table S2.**
**Name of PAMs' network topologies and analogous chemical examples.**

| Network topology of PAMs | Equivalent RCSR symbol | Chemical example |
|---|---|---|
| **J** | **nbo-b** | Niobium monoxide (*54*) |
| **T** | **pbz (hxg)** | Polybenzene (*55*) |
| **S** | **pcu-b** | Colloidal cocrystal (*56*) |
| **D** | **dia** | Diamond (*57*) |
| **C** | **pcu** | Polonium (*58*) |

**Movie S1.**
Comparison between a fused and a regular J-4-ring PAM.

**Movie S2.**
Representative compression test videos (50x speed).

**Movie S3.**
A representative simple shear test video (20x speed).

**Movie S4.**
Representative rheology test videos (20x speed).

**Movie S5.**
Simulations of spherical J-4-ring PAMs deforming under gravity.

**Movie S6.**
PAMs with S-6/2 topology but different particle geometries.

**Movie S7.**
Two stable configurations of J-4-square.

**Movie S8.**
Illustrated transition between two stable configurations of J-4-square.

**Movie S9.**
Electrostatic reconfiguration of μ-PAMs (J-4-ring) on Van de Graaff generator (real time).